\PassOptionsToPackage{table}{xcolor}
\documentclass[manuscript]{acmart}
\usepackage[english]{babel}
\usepackage{amsmath}
\usepackage{textcomp}
\usepackage{pifont}
\usepackage{siunitx}
\usepackage{tikz}
\usepackage{multirow}
\usepackage{wrapfig}
\usepackage[ruled,lined,linesnumbered,vlined,algo2e]{algorithm2e}
\usepackage{framed}
\usepackage{listings}
\usepackage{bbding}
\usepackage{enumitem}
\usepackage{threeparttable}
\usepackage{xspace}
\usepackage{comment}
\usepackage{mdframed}
\usepackage{tabularx}
\usepackage{array}
\usepackage{pgfplots}
\pgfplotsset{compat=1.18}
\usepackage{xfp}
\usepackage{makecell}
\usepackage{cleveref}
\makeatletter
\AtBeginEnvironment{algorithm2e}{%
  \let\@vspace\@vspace@orig
  \let\@vspacer\@vspacer@orig}
\AtBeginEnvironment{mdframed}{%
  \let\@vspace\@vspace@orig
  \let\@vspacer\@vspacer@orig}
\makeatother

\setcopyright{acmlicensed}
\copyrightyear{2026}
\acmYear{2026}
\acmDOI{XXXXXXX.XXXXXXX}

\acmJournal{TOSEM}
\acmVolume{}
\acmNumber{}
\acmArticle{}
\acmMonth{0}
\newcommand*\circled[2][]{\tikz[baseline=(char.base)]{
            \node[shape=circle,fill=black,draw=black,line width=0.5pt,inner sep=0.5pt, minimum size=0.5 em,#1] (char) {\textcolor{white}{#2}};}}
\newcommand{\challengeLabel}[2][\footnotesize]{{#1\sffamily\bfseries #2}}

\definecolor{janusbg}{HTML}{EAF2FB}
\definecolor{atlasbg}{HTML}{FFF2E8}
\definecolor{poirotbg}{HTML}{EAF7EF}
\definecolor{janushd}{HTML}{2C5F8A}
\definecolor{atlashd}{HTML}{B5651D}
\definecolor{poirothd}{HTML}{2E7D4F}
\definecolor{totalbg}{HTML}{E2E2E2}
\definecolor{januscr}{HTML}{BCD7F2}
\definecolor{atlascr}{HTML}{FBD3AC}
\definecolor{poirotcr}{HTML}{BFE6CE}
\newcommand{\tool}{{\textsc{Janus}}\xspace}

\newcommand{\zxy}[1]{\textcolor{black}{#1}}

\newcolumntype{C}[1]{>{\centering\arraybackslash}p{#1}}
\newcolumntype{Y}{>{\raggedright\arraybackslash}X}
\newcommand{\tabhead}[1]{\makebox[\linewidth][c]{\textbf{#1}}}
\newcommand{\bugtype}[2]{%
  \textit{#1}\par
  \noindent{\scriptsize$\hookrightarrow$~#2}%
}

\SetCommentSty{myCustomCommentFont}

\begin{document}
\title[Janus: Agentic Cross-Language Fuzzing Harness Generation for Android Native Libraries]{\tool: Agentic Cross-Language Fuzzing Harness Generation for Android Native Libraries}

\author{Xiangyu Zhang}
\affiliation{%
  \department{DISSec, NDST, College of Cryptology and Cyber Science}
  \institution{Nankai University}
  \city{Tianjin}
  \country{China}}
\affiliation{%
  \institution{Zhongguancun Academy}
  \city{Beijing}
  \country{China}}

\author{Yucheng Su}
\affiliation{%
  \institution{Xiaohongshu Inc.}
  \city{Beijing}
  \country{China}}

\author{Lingling Fan}
\correspondingauthor
\email{linglingfan@nankai.edu.cn}
\affiliation{%
  \department{DISSec, NDST, College of Cryptology and Cyber Science}
  \institution{Nankai University}
  \city{Tianjin}
  \country{China}}
\affiliation{%
  \institution{Zhongguancun Academy}
  \city{Beijing}
  \country{China}}

\author{Yinan Li}
\affiliation{%
  \department{DISSec, NDST, College of Cryptology and Cyber Science}
  \institution{Nankai University}
  \city{Tianjin}
  \country{China}}

\author{Sen Chen}
\affiliation{%
  \department{DISSec, NDST, College of Cryptology and Cyber Science}
  \institution{Nankai University}
  \city{Tianjin}
  \country{China}}
\renewcommand{\shortauthors}{Zhang et al.}

\begin{abstract} Modern Android apps extensively rely on native libraries, bridging Java/Kotlin and C/C\texttt{++} through the Java Native Interface (JNI). This cross-language interoperation, however, exposes the memory-safe Android Runtime to the unsafe C/C\texttt{++} world, creating an attack surface that has been repeatedly exploited for full device compromise. While fuzzing is a proven technique for vetting native libraries, state-of-the-art harness generation tools struggle to faithfully capture closed-source cross-language API usage: (\emph{i}) precise program analysis across the JNI boundary does not scale, whereas lightweight approximations miss critical invocation patterns; and (\emph{ii}) they fall short in constructing complex composite objects consumed by JNI methods.

In this paper, we propose \tool, the first agentic approach
to synthesizing cross-language JNI-aware
fuzzing harnesses for Android native libraries. \tool
iteratively resolves both explicit and implicit
cross-language call dependencies to recover app-specific
runtime context, and recursively assembles cross-language
argument objects in a bottom-up manner. \tool leverages
an autonomous agentic workflow to \zxy{reduce} reliance on hand-crafted
heuristics, capturing the vast landscape of hidden API
usage knowledge embedded in closed-source native libraries. Evaluation on 44 popular real-world 
apps (such as Google and YouTube Music, 
covering 21,386 JNI methods total) shows that \tool produces
over 5.9$\times$ more valid harnesses (6,094) than
state-of-the-art automatic harnessing tools. On
pre-installed apps from four mainstream Android vendors
(e.g., Xiaomi, Vivo), \tool further uncovers 6
previously unknown vulnerabilities in closed-source native
libraries. 


\end{abstract}



\maketitle

\section{Introduction}\label{sec:intro}
Native libraries are pervasive in modern Android apps, invoked from Java/Kotlin through the Java Native Interface (JNI) for performance gains or to reuse battle-tested legacy code. This convenience, however, comes at a steep security cost: the JNI bridge re-stitches the memory-safe Android Runtime (ART)
onto the unsafe C/C\texttt{++} world; once loaded, a native library shares
the hosting process's address space and executes under the same
user ID (UID) and granted permissions, yet its memory accesses are
\emph{not} protected by ART's managed-memory safety guarantees~\cite{android_memory_safety}. Memory-corruption defects in native libraries thus reintroduce a cross-language attack surface into otherwise sandboxed processes, and have repeatedly served as launchpads for full device compromise. Most recently, this attack surface has been notably weaponized
by the advanced spyware LANDFALL~\cite{unit42_landfall_2025}
and by the 0-click chain that Google Project Zero used to break
Pixel 10~\cite{google_project_zero_pixel10_2026}. Thus, \textbf{given how deeply native code is woven into today's Android ecosystem, the need for scalable approaches to vetting this fast-growing cross-language attack surface has never been more urgent.}


Android native libraries have been shown by prior work~\cite{zd, di2025hercules} to be
effectively auditable through fuzzing.
Central to this process is \emph{harness generation}, i.e., synthesizing
a wrapper program that performs the necessary initialization and
argument construction before feeding fuzzer-generated data to the
target API~\cite{sherman2026snakecharmer}. However, harnessing closed-source Android native libraries presents \textbf{two inherent issues} that have not been adequately addressed. (\emph{i})~Recovering cross-language API usage knowledge related to harness synthesis is non-trivial, as it demands knowledge scattered across
two languages. Two complementary aspects should be acquired for valid harnesses: \emph{call dependencies} (i.e., the invocation
sequence among JNIs) and \emph{argument construction} (i.e., how arguments are built and
passed)~\cite{zd,ispoglou2020fuzzgen,babic2019fudge}. (\emph{ii})~Native libraries are predominantly closed-source and privately
bound to their host apps.
Consequently, JNI usage patterns can be obtained neither from
official documentation nor from portable examples across
apps.
Each app effectively defines its own private API usage conventions,
making such knowledge largely non-transferable and requiring it to be
recovered anew for each app~\cite{di2025hercules}.

In light of this, recent studies~\cite{zd, di2025hercules} attempt to learn
JNI usage knowledge and synthesize JNI-aware harnesses
automatically. Atlas~\cite{zd} performs cross-language
symbolic execution and taint propagation in pursuit of
precise Java$\leftrightarrow$Native data-dependency
recovery, but the inherent path explosion prevents it
from scaling to contemporary commercial Android apps (only 51\%
directly usable in its 17-vendor-app dataset). Another
state-of-the-art tool, \textsc{Poirot}~\cite{di2025hercules}, takes
the opposite stance, dropping native-side analysis in
favor of lightweight Java-side data- and control-flow
analysis to mimic each app's behavior; however, the
Java-only strategy, while more scalable, sacrifices
substantial precision (only 49.5\% valid across ${\sim}$4K apps).
Both state-of-the-art artifacts share two limitations:
(\emph{i})~they struggle to capture cross-language API usage
knowledge, facing an inherent precision-scalability trade-off
in static analysis across the JNI boundary: precise cross-language
analysis is notoriously hard to scale~\cite{zd}, while lightweight approximations tend to miss
critical invocation patterns~\cite{di2025hercules}; and (\emph{ii})~they fall short of addressing how to acquire
argument-construction knowledge for the complex composite
objects routinely consumed by JNI methods.

\textbf{Root Challenges.}
At the root of the predicament (prior artifacts are trapped in) lie two fundamental challenges unique to
harnessing Android native libraries. \circled[font=\footnotesize\sffamily\bfseries,inner sep=1.2pt,minimum size=14pt]{\challengeLabel{C1}} Faithfully reconstructing cross-language invocation sequences requires intricate, app-specific runtime context (e.g., Java-side argument-object instantiation and field assignment, and native-side global state and runtime handles). This process often involves missing or misordered explicit and implicit dependencies, which can readily lead to silent failures or trivially shallow execution. 
\circled[font=\footnotesize\sffamily\bfseries,inner sep=1.2pt,minimum size=14pt]{\challengeLabel{C2}} Cross-language argument construction exposes a
non-trivial semantic gap. JNI methods in real-world
applications commonly accept complex objects (e.g.,
nested, user-defined composite types) whose construction
logic cannot be inferred from JNI type signatures alone. Since the harness executes on the native side, these Java objects require assembly from scratch
through reflection-style JNI functions~\cite{leguevel_android_greybox_fuzzing_2023}, making it inherently challenging to reconstruct objects and capture the semantic relationship that conform to native-side expectations.  

\textbf{Our Approach.} To overcome the inherent challenges above,
we propose \textbf{\tool}, an agentic approach that synthesizes
far-reaching and meaningful cross-language JNI-aware harnesses.
\tool's \textbf{key insights} lie in two algorithms interacting with
dedicated agentic workflows: To tackle C1, \tool \zxy{handles} cross-language invocation
sequences by treating unmet dependencies as the driving
force, iteratively resolving both explicit and implicit
cross-language invocation context, which spans Java-side
object instantiation and native-side runtime state. To address C2,
\tool recursively decomposes complex composite objects into directly synthesizable leaf types, then reassembles them bottom-up to construct the objects expected by the native side. \tool implements core algorithms within an agentic
workflow powered by Large Language Models (LLMs), capturing the JNI usage knowledge and semantic relationships between parameters, reducing reliance on brittle hand-crafted heuristics, thereby empirically mitigating the
precision-scalability trade-off faced by prior work.

\textbf{Agentic Design.} \tool builds an agentic workflow tailored for JNI-aware harness synthesis, in which the underlying LLMs act as autonomous agents that perform
code reasoning, planning, tool invocation, and self-reflection
throughout harness synthesis, rather than serving as auxiliary
components in a conventional fuzzing pipeline~\cite{huang2024understanding,zeng2025pbfuzz}. This agentic design, however, surfaces three specific challenges:
(\emph{i}) aimless exploration from sprawling cross-language code fragments;
(\emph{ii}) app-specific runtime context inconsistency over long-horizon
reasoning; and (\emph{iii}) hallucinations in JNI-aware harness generation.
To sidestep the trap of overwhelming the LLM with sprawling
cross-language code fragments and asking it to single out the
critical pieces, \tool equips the agent with a tool-augmented,
structure-aware code retriever that forestalls aimless exploration.
To tame context drift and hallucinations, \tool persists verified
intermediate artifacts across cross-language methods in a dedicated
memory layer and closes the generation loop with a validation-guided
iterative refinement that attributes failures to the responsible
synthesis stage for targeted refinement.

\textbf{Results.} We evaluate \tool on 21{,}386 JNI methods of 44 real-world
Android apps against two state-of-the-art
harnessing tools for Android native libraries, Atlas++~\cite{atlasplusplus}
and \textsc{Poirot}~\cite{di2025hercules}.
\tool produces $7.4\times$/$5.9\times$ more valid harnesses, and achieves
$5.8\times$/$6.5\times$ the bitmap coverage compared with
Atlas++ and \textsc{Poirot}, respectively.
These results suggest that \tool can exercise more and deeper
native code functionalities, exposing more potential security vulnerabilities
and defects. \tool further uncovers \zxy{6} previously
unknown memory-corruption vulnerabilities in closed-source native
libraries shipped by mainstream Android vendors
(e.g., Xiaomi, Vivo).


This paper makes the following three contributions:
\begin{itemize}
  \item We propose \tool, the first agentic harness-synthesis approach toward JNI-aware fuzzing of Android native libraries, \zxy{paving the way for broader auto-harnessing techniques across the Java-native boundary.}
    \item We present algorithms to guide agentic workflows for 
    automated harness generation that resolve explicit and implicit 
    call dependencies and recursively construct composite arguments, 
    backed by structure-aware retrieval, validation-guided refinement, 
    and harness memory.
  \item \tool outperforms the state of the art in harness synthesis, 
    execution validity, and code coverage, and further uncovers previously unknown vulnerabilities in closed-source native libraries shipped with mainstream Android vendor apps. For future research, we have released our prototype and artifacts on GitHub~\cite{janus_anonymous_repo}.
\end{itemize}

\section{PRELIMINARIES}

\subsection{Java Native Interface in Android}

In Android, the Java Native Interface (JNI) lets apps declare native methods with the \texttt{native} keyword and provide a matching C/C\texttt{++} implementation. Methods are bound either statically through naming conventions resolved by \texttt{dlsym} at load time, or dynamically via \texttt{JNI\_OnLoad} calling \texttt{RegisterNatives}~\cite{android_jni_tips}. Once bound, each native call receives a \texttt{JNIEnv*} pointer, an opaque handle that mediates all interaction with the managed heap. Because native code cannot directly access Java object layouts, the JNI specification exposes a reflective API through \texttt{JNIEnv}: reading a field requires consecutive calls to \texttt{FindClass}, \texttt{GetFieldID}, and \texttt{Get<Type>Field}; object creation follows a similar sequence of \texttt{FindClass}, \texttt{GetMethodID} (for \texttt{<init>}), and \texttt{NewObject}~\cite{wei2018jn}. For nested or user defined composite types, these reflective lookups should be chained layer by layer, as illustrated in Fig.~\ref{figmoti} (lines N9-N16).

\subsection{\zxy{LLM Agents for JNI-Aware Harness Generation}}
\label{sec:llm-agents}

LLM agents, including OpenCode~\cite{opencode_github}, Claude Code~\cite{anthropic_claude_code}, and OpenAI Codex~\cite{openaicodex}, combine language-model reasoning with external tools to inspect program artifacts and execute generated code~\cite{huang2024understanding,zeng2025pbfuzz,zhang2024effective}. Operating in an iterative loop of code reasoning, planning, tool
invocation, and self-reflection, the agent identifies missing
information, queries relevant code, and revises its output based
on execution observations~\cite{yao2023react,shinn2023reflexion}.
For JNI-aware harness generation, the required program context 
spans Java-side call sites and constructors as well as native-side
implementations; by inspecting these artifacts, the agent
investigates which invocations establish the required runtime
state and how argument objects should be initialized, thereby
connecting app-specific API usage knowledge across the language
boundary.

Within a limited context window, the agent needs to acquire new cross-language code facts while retaining relevant intermediate findings~\cite{packer2023memgpt}. To organize such multi-step tasks, OpenCode supports Skills that encode reusable procedural instructions and are loaded into the agent's context on demand~\cite{opencode_skills}. These Skills specify task decomposition, intermediate outputs, decision criteria, and validation and termination conditions, while tools provide executable operations for code retrieval, compilation, and runtime testing~\cite{opencode_tools}. \tool adopts OpenCode because its modifiable open-source implementation and support for custom tools and Skills facilitate integrating the JNI-aware code retriever with the algorithm-guided harness synthesis and refinement workflow, while also promoting research reproducibility.

\subsection{Motivating Example and Challenges}

\begin{figure}
    \centering
    \includegraphics[width=1 \textwidth]{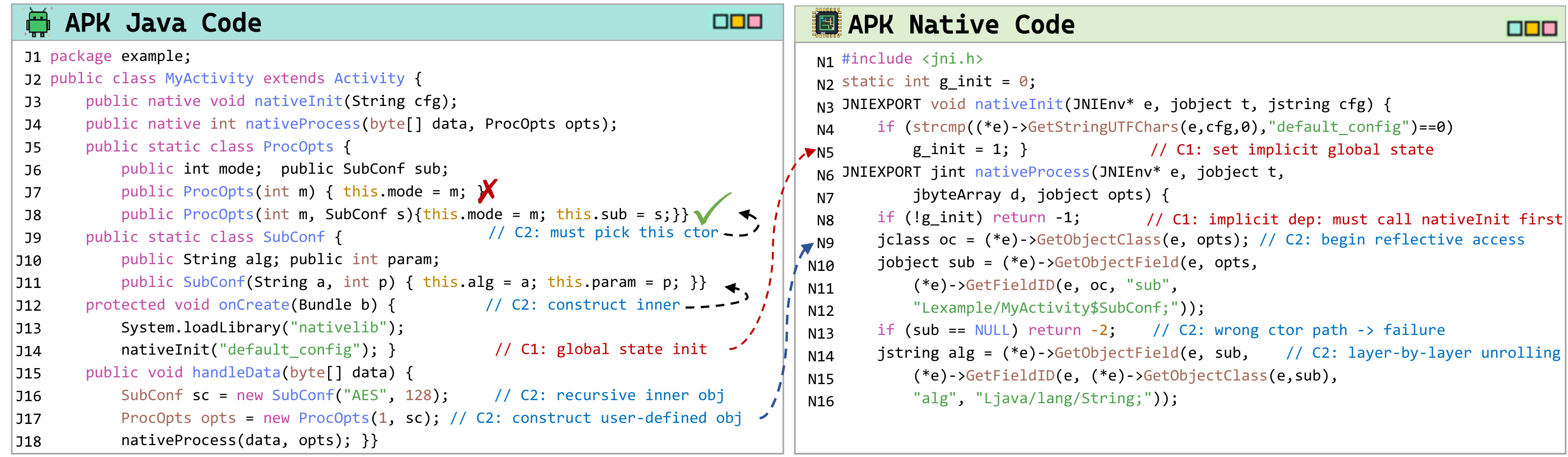}
    \caption{{Motivating case.}}
    \label{figmoti}
\end{figure}

We motivate \tool by exposing two challenges \emph{inherent} to JNI-aware harness generation (\S\ref{c1}, \S\ref{c2}), using a concrete example (Fig.~\ref{figmoti}). \S\ref{c3} then shows, through a preliminary study, that naively delegating the task to an LLM agent fails to resolve these obstacles and instead introduces a third, \emph{solution-induced} challenge.

Specifically, during normal execution of the real-world decompiled app in Fig.~\ref{figmoti} (spanning decompiled Java lines~J1--J18 and native lines~N1--N16), \texttt{MyActivity.onCreate} loads \texttt{nativelib} and invokes \texttt{nativeInit} (J3) with a configuration string to initialize the internal library state (J12--J14, N3--N5); subsequently, \texttt{handleData} instantiates the nested user-defined classes \texttt{SubConf} (J9--J11) and \texttt{ProcOpts} (J5--J8) to invoke the target JNI method \texttt{nativeProcess} (J4, J15--J18), which verifies this state and reflectively retrieves nested fields from \texttt{opts} before reaching its core data-processing logic (N6--N16).
However, synthesizing an effective fuzzing harness that achieves \textbf{{valid execution}} of a target JNI method proves to be challenging.
Harnessing closed-source Android native libraries inherently presents difficulties because the prerequisites governing valid execution are intertwined across the Java and C/C++ boundary.
Furthermore, meaningfully exercising the target JNI method not only requires recovering the cross-language runtime context, spanning prior native state initialization and Java-side setup (\S\ref{c1}), but also demands synthesizing nested composite argument objects whose internal field requirements are implicitly dictated by the native implementation (\S\ref{c2}).
Below, we delve into each of these obstacles in detail.

\subsubsection{\textbf{Hidden App-Specific Invocation Context}}
\label{c1}

JNI method invocations typically depend on app-specific context established by preceding Java-side and native-side operations. Consider \texttt{nativeProcess} in Fig.~\ref{figmoti}: its valid execution depends on two pieces of runtime context. (\emph{i}) The \texttt{ProcOpts} object passed from the Java side needs to be initialized through the constructor \texttt{ProcOpts(int, SubConf)} (J8) so that its nested field \texttt{sub} is non-null (J6). (\emph{ii}) The native-side global flag \texttt{g\_init} must have been set to~1 by the prior JNI call \texttt{nativeInit} (N5). If either condition is missing, \texttt{nativeProcess} returns an error code at its early guard checks (N8 or N13) and never reaches the core logic. Existing work attempts to recover such calling context
through static analysis, yet with limited effectiveness. For instance, \textsc{Poirot} abandons native-side analysis entirely and extracts call
sequences solely from the Java side: expanding outward from the
callsite \texttt{handleData} of \texttt{nativeProcess}, it obtains
only \texttt{nativeProcess} itself, while the critical
\texttt{nativeInit} resides in a separate entry point
\texttt{onCreate} that lies outside the call chain and is therefore
omitted.

\subsubsection{\textbf{Missing Object-Construction Semantics}}
\label{c2}

In real-world applications, JNI methods routinely accept complex objects \zxy{(e.g., user-defined composite objects such as \texttt{ProcOpts} and its inner \texttt{SubConf} in~Fig.~\ref{figmoti}, lines~J4--J11)} whose construction needs to be unrolled layer by layer across the language boundary.
\zxy{Because the fuzzing harness executes on the native side where Java object parameters are exposed as opaque references (e.g., \texttt{jobject opts} at line~N7), the harness cannot directly populate object memory layouts; instead, it must assemble these Java objects from scratch via reflection-style JNI APIs (e.g., \texttt{FindClass}, \texttt{GetMethodID}, and \texttt{NewObject}).
However, how a composite object should be constructed cannot be inferred from type signatures or Java class definitions in isolation, as it requires aligning Java-side instantiation paths with the fields internally accessed by the native implementation.}
Specifically, in~Fig.~\ref{figmoti}, \texttt{nativeProcess} reads \texttt{ProcOpts.sub.alg} step by step through a chain of JNI reflective calls (\zxy{lines~N9--N16}): \zxy{it first reflectively retrieves \texttt{opts.sub} (aborting if \texttt{sub == NULL} at lines~N10--N13) and subsequently reads \texttt{sub.alg} via \texttt{GetFieldID}/\texttt{GetObjectField} (lines~N14--N16).}
\zxy{Satisfying these native-side field accesses} requires the harness not only to select the correct constructor \texttt{ProcOpts(int, SubConf)} \zxy{(line~J8) over \texttt{ProcOpts(int)} (line~J7)}, but also to recursively construct the inner \texttt{SubConf} object \zxy{via \texttt{SubConf(String, int)} (line~J11)} and supply a legal value for its \texttt{alg} string field.
Existing methods~\cite{zd, di2025hercules} are powerless in this regard, as both of them largely neglect to provide a means to synthesize complex Java objects, making it difficult to achieve a valid execution of a target JNI method and leaving JNI methods that consume composite types largely untestable.

\subsubsection{\textbf{LLM Agents Are Not a Silver Bullet}}
\label{c3}

\begingroup
\zxy{Given that overcoming both challenges demands reasoning over scattered cross-language code semantics rather than relying on rigid static heuristics, a natural question arises: \textit{can modern LLM agents (\S\ref{sec:llm-agents}) directly bridge this gap by autonomously inspecting APK artifacts and synthesizing JNI-aware harnesses?} To answer this question, we conducted an exploratory study using a state-of-the-art general-purpose coding agent.}

We randomly sampled 10 Android application packages (APKs), comprising 1,450 JNI methods in total, from the dataset of \textsc{Poirot}~\cite{di2025hercules}. We instantiated the general-purpose LLM-based agent using Claude Code~\cite{anthropic_claude_code}, with Claude Opus 4.5~\cite{anthropic_claude_opus_4_5_2025}, one of the most capable publicly available models \zxy{leading major coding and agentic software engineering leaderboards (e.g., SWE-bench and LiveCodeBench)} at the time of this study~\cite{jimenez2024swebench, jain2025livecodebench, chiang2024chatbot}, as the underlying model. We provided the agent with basic reverse-engineering tools, including JADX \cite{jadx} and Ghidra \cite{ghidra}, as well as an Android emulator. We imposed no budget limit on this experiment. To keep the input conditions consistent, the agent received only APK files as input artifacts, the same input type as the full \tool implementation. Fig.~\ref{fig:general-agent-prompt} presents an expanded prompt template for this task.

\begin{figure}[t]
\centering
\zxy{%
\begin{tikzpicture}[
  promptpanel/.style={
    anchor=north west,
    text width=\dimexpr0.8\columnwidth-16pt\relax,
    inner xsep=8pt,
    inner ysep=4pt,
    outer sep=0pt,
    align=left,
    font=\footnotesize,
    fill=black!5
  }
]
\node[promptpanel] (role) at (0,0) {%
  \textbf{Role:} You are a software engineering expert responsible for generating fuzzing harnesses for JNI methods in closed-source Android apps.};
\node[promptpanel] (objective) at (role.south west) {%
  \textbf{Objective:} Given the target app's APK, generate JNI-aware fuzzing harnesses that exercise its native methods using the Android Runtime (ART) in an Android emulator.};
\node[promptpanel] (input) at (objective.south west) {%
  \textbf{Input:} APK file of the target application: \texttt{\{target\_apk\}}.};
\node[promptpanel] (tools) at (input.south west) {%
  \textbf{Available Tools:} JADX and Ghidra for inspecting decompiled Java and native code, together with an Android emulator for executing the generated harnesses.};
\node[promptpanel] (requirements) at (tools.south west) {%
  \textbf{Task Requirements:}
  \begin{enumerate}[nosep,topsep=2pt,leftmargin=1.5em,label=\arabic*.]
    \item Inspect the APK to identify its JNI methods and their corresponding native implementations.
    \item Generate C/C\texttt{++} harness code that loads the target native libraries and invokes the JNI methods with fuzzer-controlled inputs.
    \item Provide the runtime initialization and argument preparation required to invoke the target methods in ART.
    \item Attempt to build and execute the generated harnesses in the emulator. Use the observed errors to revise the code, and report any unresolved issues.
  \end{enumerate}
};
\node[promptpanel] (output) at (requirements.south west) {%
  \textbf{Expected Output:} Return the harness source code and the instructions required to build and run it. Identify the target JNI methods and report whether the harnesses were successfully built and executed.};
\node[
  anchor=south west,
  text width=\dimexpr0.8\columnwidth-16pt\relax,
  inner xsep=8pt,
  inner ysep=5pt,
  outer sep=0pt,
  align=center,
  font=\small,
  fill=black!65,
  text=white,
  rounded corners=3pt
] (heading) at (role.north west) {Prompt for the General-Purpose LLM Agent};
\foreach \panelname in {role,objective,input,tools,requirements}{%
  \draw[black!55,densely dashed,line width=0.5pt]
    (\panelname.south west) -- (\panelname.south east);
}
\draw[black!65,line width=0.6pt,rounded corners=3pt]
  (heading.north west) rectangle (output.south east);
\end{tikzpicture}%
}
\caption{\zxy{Prompt template for the general-purpose LLM agent.}}
\label{fig:general-agent-prompt}
\Description{Prompt template giving an LLM agent its role, objective, target APK, available tools, task requirements, and expected harness outputs.}
\end{figure}

\zxy{Two authors cross-validated the generated results for all 1,450 JNI methods, confirming that the off-the-shelf agent generated 45 buildable and 13 runnable harnesses, which involved simple JNI functions with basic parameters. By inspecting its execution traces, we identified three compounding
failure modes, which jointly indicate that the unconstrained
reasoning of an off-the-shelf, general-purpose LLM agent is
insufficient for the app-specific cross-language semantic
reasoning demanded by \S\ref{c1} and \S\ref{c2}.}
\endgroup

(\emph{i})~\textbf{Aimless exploration.}
Decompiled real-world APKs can reach hundreds of thousands of lines,
with code facts scattered across both the Java and C/C\texttt{++} sides.
At this scale, the agent either sees only local slices due to
limited context capacity, or wanders through irrelevant files and
conflates syntactically similar but semantically unrelated
fragments, building its reasoning on fragmented and unverifiable
snippets.
(\emph{ii})~\textbf{Inconsistency over long horizons.}
Harnessing a single JNI method requires dozens of reasoning steps,
and an app often contains hundreds of targets sharing parameter
types and precursor dependencies. Without coordination, the agent
repeatedly unfolds the same sub-problems, quickly exhausting the
context window; as the dialog grows, attention to earlier facts
decays, leading to self-contradictory call sequences or forgotten
construction procedures.
(\emph{iii})~\textbf{Hallucination without validity refinement.}
Even when equipped with basic compilation and execution tools, the agent fabricates non-existent entities (e.g., fictitious field names,
constructors, or JNI return-value semantics) and relies on them in
subsequent reasoning. More subtly, it interprets ``no crash'' as
success, while execution may complete only shallow steps without
reaching the core logic, yielding no effective coverage and masking
real defects.

\section{Approach}

\begin{figure}
    \centering
    \includegraphics[width=\textwidth]{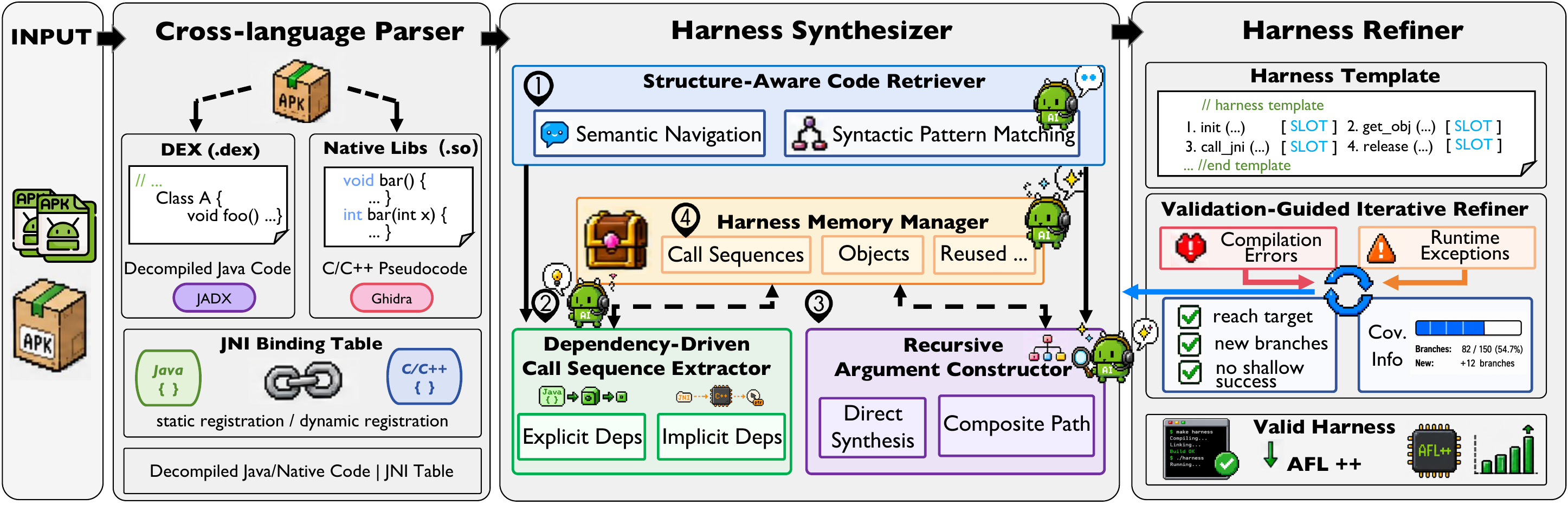}
    \caption{Overview of \tool.}
    \label{fig:overview}
\Description{Overview of Janus, showing APK preprocessing, harness synthesis with code retrieval, call-sequence extraction, recursive argument construction and memory, followed by runtime refinement.}
\end{figure}

To address the challenges of generating harnesses for Android native libraries, we introduce \tool, a tailored agentic approach for cross-language JNI-aware harness generation. \tool unfolds in three key stages, as shown in Fig.~\ref{fig:overview}. \tool takes an APK file as input and employs a divide-and-conquer strategy, separating it into DEX bytecode (\texttt{.dex}) and native libraries (\texttt{.so}). Specifically, \circled[font=\fontsize{9}{10}\selectfont\sffamily\bfseries,inner sep=1pt,minimum size=12pt]{\challengeLabel[\fontsize{9}{10}\selectfont]{1}}~\emph{Cross-Language Parser} grounds downstream harness synthesis by recovering a unified view of Java–native interactions, dispelling the opacity introduced by JNI indirection. \circled[font=\fontsize{9}{10}\selectfont\sffamily\bfseries,inner sep=1pt,minimum size=12pt]{\challengeLabel[\fontsize{9}{10}\selectfont]{2}}~\emph{Harness Synthesizer}, built on a structure-aware retrieval layer, iteratively recovers each JNI method's precursor call sequence, uncovering both explicit and implicit dependencies, and recursively assembling its argument objects in a bottom-up manner; in parallel, the harness memory persists verified intermediate artifacts across methods, alleviating the context pressure of long-horizon reasoning. \circled[font=\fontsize{9}{10}\selectfont\sffamily\bfseries,inner sep=1pt,minimum size=12pt]{\challengeLabel[\fontsize{9}{10}\selectfont]{3}}~\emph{Feedback-Driven Harness Refiner} constrains the generation space with app-specific harness templates and refinement, attributing compilation errors, runtime exceptions, and coverage signals back to the upstream stage responsible for repair, thereby suppressing the hallucinations and omissions that accumulate in long-chain reasoning. 

\subsection{Phase 1: Cross-Language Parser}
\label{phase1}

Collectively, the cross-language analyses provide \tool with the
initial cross-language context needed to trace how the consumer application
invokes and propagates objects into the native library, paving
the way for subsequent harness synthesis that mirrors realistic
API usage patterns~\cite{zhang2024effective}.
On the Java side, \tool extracts structural information for each
defined class and method, such as fully qualified signatures,
owning classes, argument and return types, and inter-method call
relationships. Following Java-side parsing, \tool analyzes the
native libraries and extracts function bodies along with the
implicit constraints they impose on incoming arguments (e.g.,
validity checks on file descriptors or object fields). 

To further
bridge the two sides, \tool resolves the runtime address of every
native method on-device through the ART, covering both
static registration via \texttt{dlsym} and dynamic registration via
\texttt{RegisterNatives} hooking~\cite{android_jni_tips}. The resulting \emph{JNI binding
table} records, per method, its fully qualified Java signature, argument and return types, owning \texttt{.so}, and relative offset. Cross-language analyses ground \tool's understanding in a unified view of Java–native interaction behavior, thereby 
laying the foundation for synthesizing harnesses that systematically 
exercise JNI functionality.

\subsection{Phase 2: Harness Synthesizer}
\label{phase2}

The Harness Synthesizer, the core stage of \tool,
synthesizes JNI-aware harnesses designed to faithfully exercise app-specific
Java-Native interactions, equipped with the cross-language decompiled artifacts (\S\ref{phase1}).
This stage comprises four components. \textit{Structure-Aware Code Retriever} equips \tool with cross-language code retrieval and comprehension capabilities. Building atop this, the \emph{Dependency-Driven Call Sequence Extractor} addresses the question of \emph{how a JNI method should be invoked}, producing JNI method cross-language call sequences; the \emph{Recursive Argument Constructor} then operates on these sequences to resolve \emph{how arguments should be constructed}, producing argument construction schemes. Spanning the entire stage, the \emph{Harness Memory Manager} persists verified intermediate artifacts, enabling \tool to reuse validated components across the JNI boundary, thereby mitigating redundant exploration and re-verification and alleviating the context pressure inherent in long-horizon tasks.

\subsubsection{\textbf{Structure-Aware Code Retriever}}
\label{sec:phase2.1CodeRetriever}
The decompiled artifacts from the Cross-Language Parser are voluminous, with harness-relevant facts scattered across decompiled Java code and native pseudocode. Textual retrieval alone struggles to reliably capture such cross-language associations, causing critical facts to be lost during the reasoning pass of LLM agents and thereby degrading the quality of the synthesized harness. To address this, \tool{} builds a structure-aware retrieval layer atop the decompiled code, allowing the underlying agent to query along the symbol relationships and syntactic structures inherent in the code itself.


\noindent $\bullet$ \textbf{Semantic Navigation.}
Existing off-the-shelf language servers (LSs), namely
\textit{jdtls}~\cite{eclipse_jdt_ls_github} for Java and
\textit{clangd}~\cite{clangd} for native code, implement the
Language Server Protocol (LSP), but each is confined to a single
language and cannot support bidirectional Java$\leftrightarrow$native
navigation. To bridge this gap, \tool
implements a JNI-aware cross-language LS that augments the two
underlying servers with the JNI method table (\S\ref{phase1}),
thereby linking their language-specific symbol indexes. For a cross-boundary query, such as \textit{Java declaration} $\rightarrow$ \textit{native implementation} or the reverse, our server resolves the mapping, returning the target location together with its owning library and relative offset. Once the query reaches either side, the
cross-language LS delegates intra-language reference, definition,
and type queries to the corresponding underlying LS. This design
turns the two otherwise disjoint symbol indexes into a unified,
traversable cross-language symbol table.
 
\noindent $\bullet$ \textbf{Syntactic Pattern Matching.}
Resolving JNI behavior requires identifying structural code
patterns, such as native functions that read a receiver field
via \texttt{GetFieldID}, which cannot be expressed as symbol
queries. \tool therefore defines two sets of AST-Grep templates:
Java-side templates capture declarations of native functions and
object-propagation paths (e.g., constructors, setters, field
read/writes), while native-side templates capture JNI field and
object accesses (e.g., \texttt{GetFieldID}). All matches are
returned uniformly as $\langle \textit{location},\allowbreak \textit{symbol
info},\allowbreak \textit{code snippet} \rangle$ triples.


\tool adaptively selects among these tools as the LLM agent's reasoning proceeds,
so that each synthesis decision is traceable to a specific source
location, equipping the subsequent phase with the unified
cross-language view required to generate far-reaching JNI-aware
harnesses.

\subsubsection{{\textbf{Dependency-Driven Call Sequence Extractor}}}
\label{phase2.2call-sequence-extractor}

\tool recovers the app-specific runtime context of prerequisite calls for each target JNI method, enabling the harness to exercise execution paths legitimately reachable by the consumer application~\cite{di2025hercules}. However, due to the closed-source nature of Android native libraries, JNI usage knowledge is often sparse (or even missing), making app-specific runtime context recovery inherently challenging: \textbf{incomplete or misordered prerequisite calls may silently corrupt the execution environment, inducing spurious crashes}~\cite{zd}. The key, therefore, is for the harness to faithfully reconstruct the prerequisite context on which the target JNI method depends during legitimate execution in a real-world application. By examining the execution constraints of target JNI methods in real apps, we identify two categories of dependency that should be satisfied:

\noindent $\bullet$ \textbf{Explicit Dependencies.} Constraints that can be established entirely on the Java side. Concretely, an argument object needs to be instantiated through a Java constructor, and any of its fields that the native code will subsequently read must have been explicitly assigned by Java code. Satisfying such constraints involves only Java-side object construction and field writes, without crossing the JNI boundary.

\noindent $\bullet$ \textbf{Implicit Dependencies.} Constraints that should be established on the native side by preceding JNI calls. They manifest as fields on an argument object written back from native code (e.g., C pointers stored as \texttt{long}), global context maintained inside a native library, or argument values consumed by a later call but produced as the return value of an earlier one. Their common feature is that the relevant values or state cannot be readily fabricated on the Java side and must be produced by some preceding JNI call during native execution~\cite{zd}.

The dependencies are app-specific and intertwined across
the Java and native sides, making them challenging to
acquire for reconstructing the invocation context, where
static analysis faces an inherent precision-scalability
trade-off. \tool therefore casts this recovery as a
dependency-driven algorithm that treats unmet dependencies
as the driving force, iteratively resolving both explicit
and implicit dependencies needed to reach the target JNI
method. Within this algorithm, \tool replaces handcrafted
heuristics with LLM agents that capture diverse
app-specific conventions, identifying missing dependencies
and the calls that satisfy them.


We formalize the above solving process as follows. Given a target JNI method $m$, its prerequisite call sequence is denoted $S_m = \langle c_1, c_2, \ldots, c_k, m \rangle$, where each $c_i$ is a Java method or JNI method that contributes to establishing the execution context of $m$. Every prerequisite call in the sequence is paired with a structured annotation $a_i = \langle c_i, \mathit{role}_i, \mathit{source}_i, \mathit{binding}_i \rangle$: $\mathit{role}_i$ identifies the dependency category that $c_i$ satisfies, \zxy{$\mathit{source}_i$ records the JNI-related code supporting its inclusion}, and $\mathit{binding}_i$ describes how the products of $c_i$ are consumed downstream: \emph{(i)} how its return value flows into the argument of a subsequent call and \emph{(ii)} how its side effects (e.g., writes to object fields, initialization of global context) establish a particular dependency state for later calls or for the target JNI itself. The set of currently unsatisfied dependencies of $m$ is denoted $D_m = \{d_1, d_2, \ldots, d_n\}$, where each $d_j$ corresponds to an unsatisfied instance of one of the dependency categories introduced above. The objective is to construct $S_m$ together with its annotation set $A_m$ such that $D_m \subseteq \mathrm{Resolution}(S_m, A_m)$, where $\mathrm{Resolution}(S_m, A_m)$ denotes the set of dependencies that the current sequence and its annotations are able to establish. The recovered invocation context is considered \emph{resolved} once currently identified dependencies $D_m$ are satisfied.



\begin{algorithm2e}[t]
\setcounter{AlgoLine}{0}
\caption{Call Sequence Extraction}
\label{algo1}
\DontPrintSemicolon
\SetAlgoVlined
\SetInd{0.5em}{0.5em}
\setlength{\algomargin}{0.5em}
\small
\SetCommentSty{mycommfont} 
\KwIn{$m$: target JNI method; $\Gamma$: decompiled code; $\mathcal{T}$: tool set}
\KwOut{$S_m$: accepted prerequisite call sequence; \mbox{$A_m$: annotation set, or \textsc{Fail}}}
$S_m \gets \langle m \rangle$; $A_m \gets \emptyset$ 
$D_m \gets \text{InitialDependencies}(m, \Gamma, \mathcal{T})$\;
\While{$D_m \neq \emptyset$ \textbf{and} budget remains}{
    $Q \gets \text{Agent.GenerateQueries}(m, S_m, D_m)$\;
    $E \gets \text{ExecuteTools}(Q, \mathcal{T})$\;
    $C \gets \text{Agent.ProposeCalls}(E, D_m)$\;
    $\langle S_m, A_m \rangle \gets \text{InsertAndAnnotate}(S_m, A_m, C, E)$\;
    $result \gets \text{Agent.CheckResolution}(m, S_m, A_m, D_m)$\;
    \uIf{$result == \textsc{RESOLVED}$}{
        $D_m \gets \emptyset$; 
    }
    \Else{
        $D_m \gets result.\text{missing\_dependencies}$\;
    }
}
\lIf{$D_m \neq \emptyset$}{\Return{\textsc{Fail}}}
$verdict \gets \text{Agent.GlobalReview}(m, S_m, A_m)$\;
\uIf{$verdict == \textsc{Accept}$}{\Return{$S_m, A_m$}}
\lElse{\Return{\textsc{Fail}}}
\end{algorithm2e}

\tool guides this resolution process through an agentic 
iterative loop (Algorithm~\ref{algo1}). The input consists of the target JNI method $m$ and the  \textit{Code Retriever} toolset $\mathcal{T}$; the output is an accepted prerequisite call sequence $S_m$ paired with its structured annotations $A_m$, or ${fail}$ if no such sequence can be assembled within the search budget. Specifically, $S_m$ is first initialized to contain only the target JNI method itself, $A_m$ is initialized to empty, and the initial dependency set $D_m$ is extracted from $m$'s Java declaration and native implementation (line~1). The sequence and its annotations are then gradually augmented as $D_m$ is progressively resolved during the loop.

In each round, \tool generates targeted queries from the currently unsatisfied dependencies (line~3), invokes $\mathcal{T}$ to retrieve JNI-related code (line~4), turns the retrieved results into candidate calls (line~5), and inserts them into the current sequence together with their structured annotations (i.e., ${role}$, ${source}$, ${binding}$, line~6). After insertion, \tool switches to an auditor's perspective and checks whether the sequence has been locally resolved, with the remaining unsatisfied dependencies (or $\emptyset$ upon resolution) carried into the next round ((lines 7–11).
If the search budget is exhausted before closure, the algorithm returns failure (line~12).
At the start of each round \tool first consults the memory manager (\S\ref{phase2.4harness-memory}): if a previously recorded sequence already covers a current dependency, it is reused directly to avoid redundant solving.
Once $D_m$ is satisfied, \tool performs a global review over the entire sequence (line~13), checking that the call ordering, dependency consistency, and binding relations indeed hold at the sequence level; only sequences that pass this review are finally accepted (lines~14--15).

Through this loop, \tool reframes call sequence construction
into a code-grounded resolution process driven by missing
dependencies, with the global review checking
sequence-level consistency. The output is the set of
calls (together with their supporting code) required to
reconstruct the target JNI's execution context, which serves as
the input to the subsequent argument synthesis stage.

\subsubsection{\textbf{Recursive Argument Constructor}}
\label{phase2.3arg-obj-constructor}

Producing executable harnesses further requires assembling
argument objects from scratch via reflection-style JNI
functions~\cite{leguevel_android_greybox_fuzzing_2023}, a
non-trivial task since JNI methods commonly accept complex
composite types whose construction logic cannot be inferred
from type signatures alone: \textbf{failing to align Java-side instantiation paths with the fields internally accessed by the native implementation leaves inner objects unpopulated, causing the native method to abort without achieving a valid execution}.

To tackle this challenge, \tool recursively unrolls the construction of composite objects layer by layer, selecting appropriate Java constructors and populating nested fields to satisfy native-side accesses. Specifically, Algorithm~\ref{algo:RecArgConstruction} takes as input the target argument type $\tau$, the call sequence $S_m$ \zxy{and its annotations $A_m$} produced in~\S\ref{phase2.2call-sequence-extractor}, and the tool set $\mathcal{T}$ from~\S\ref{sec:phase2.1CodeRetriever}; it returns either the construction code $\mathit{ctor}(\tau)$, which can be embedded into the harness, or \textit{fail} when no feasible candidate path exists.

\begin{algorithm2e}[tbp]
\setcounter{AlgoLine}{0}
\caption{Recursive Argument Object Construction}
\label{algo:RecArgConstruction}
\DontPrintSemicolon
\SetAlgoVlined
\SetInd{0.5em}{0.5em}
\setlength{\algomargin}{0.5em}
\small
\SetCommentSty{mycommfont}
\KwIn{$\tau$: target argument type; $S_m, A_m$: call sequence and annotations (\S\ref{phase2.2call-sequence-extractor});\;
      $\Gamma$: decompiled code; $\mathcal{T}$: tool set}
\KwOut{$\textit{ctor}(\tau)$: construction code that produces an argument of type $\tau$, or \textsc{Fail}}

\tcp{\textbf{Direct argument resolution (binding, synthesis, and reuse)}}
\lIf{$\text{Bound}(\tau, A_m)$}{\Return{$\text{EmitBoundValue}(\tau, A_m)$}}
\lIf{$\text{Synthesizable}(\tau)$}{\Return{$\text{EmitJNISynthesis}(\tau)$}}
\lIf{$\text{Memory.Hit}(\tau, S_m, A_m)$}{\Return{$\text{Memory.Get}(\tau, S_m, A_m)$}}

\tcp{\textbf{Composite object path selection}}
$\textit{ctorCands} \gets \text{QueryConstructorsAndFactories}(\tau, \Gamma, \mathcal{T})$\;
\While{$\textit{ctorCands} \neq \emptyset$}{
    $\pi \gets \text{Agent.SelectConstructionPath}(\textit{ctorCands}, S_m, A_m, \Gamma)$\;
    $\textit{ctorCands} \gets \textit{ctorCands} \setminus \{\pi\}$; $\textit{argCodes} \gets \emptyset$\;

    \tcp{\textbf{Sub-argument recursion and assembly}}
    \ForEach{$\tau' \in \text{ParamTypes}(\pi)$}{
        $\textit{ctor}(\tau') \gets \text{RecConstruct}(\tau', S_m, A_m, \Gamma, \mathcal{T})$\;
        \lIf{$\textit{ctor}(\tau') = \textsc{Fail}$}{\textbf{break to next candidate in Line 5}}
        $\textit{argCodes}.\text{add}(\textit{ctor}(\tau'))$\;
    }
    $\textit{ctor}(\tau) \gets \text{EmitNewObject}(\pi, \textit{argCodes})$\;

    \tcp{\textbf{Local Check and memoization}}
    \If{$\text{LocalCheck}(\textit{ctor}(\tau))$}{
        $\text{Memory.Put}(\tau, S_m, A_m, \textit{ctor}(\tau))$\;
        \Return{$\textit{ctor}(\tau)$}\;
    }
}
\Return{\textsc{Fail}}\;
\end{algorithm2e}

\noindent $\bullet$ \textbf{\zxy{Direct Argument Resolution.}}
This step defines the termination condition (i.e., leaf case) of the recursion:
before attempting any construction path, \textsc{Janus} first
checks whether the target type $\tau$ can be produced directly,
terminating the recursion immediately if \zxy{any of three} conditions
holds (\zxy{lines 1--3}). \zxy{(\emph{i})~$\tau$ is already bound in $A_m$ (e.g., a fixed setup literal such as \texttt{"default\_config"} in~Fig.~\ref{figmoti} or a native handle returned by an earlier call in $S_m$), in which case \textsc{Janus} directly emits the bound constant or variable reference. (\emph{ii})~$\tau$ is an unbound leaf type that} is \emph{directly synthesizable}, i.e.,
it requires no nested object assembly. This covers \ding{192}~primitive types,
obtained by slicing the corresponding number of bytes from the fuzzer
stream and decoding them (e.g., taking 4 bytes and decoding them as a
\texttt{jint}), and \ding{193}~reference types such as \texttt{String} or
primitive arrays, parsed from bytes and materialized as Java
objects via the corresponding JNI API (e.g., \texttt{NewStringUTF}).
(\emph{\zxy{iii}})~A compatible construction for $\tau$ \zxy{that covers the fields and bindings required by $(S_m, A_m)$} has already been recorded in the harnessing memory (\S\ref{phase2.4harness-memory}), in which case \textsc{Janus} reuses it directly, short-circuiting the recursion to eliminate redundant reasoning.

\noindent $\bullet$ \textbf{Composite Object Path Selection.} When $\tau$ cannot be synthesized directly, \tool enumerates candidate construction paths ${ctorCands}$, including Java constructors, factory methods, and instantiation sites of $\tau$ observed in the app (\zxy{line 4}), based on the structure-aware retrieval tools $\mathcal{T}$. This step is difficult for purely heuristic filling: a given type often exposes multiple constructors or factory methods that differ in their initialization scope, side effects, and prerequisite dependencies, and these differences are hard to tell apart from type signatures alone. \tool therefore guides the LLM agent's decision-making via
explicit task specifications that weigh two factors:
(\emph{i}) how the type is instantiated in the app, and
(\emph{ii}) how well each candidate covers the fields read
on the native side, to select the path ($\pi$) that 
matches the object state expected by the native code (\zxy{line 6}). For example, for \texttt{ProcOpts} in~Fig.~\ref{figmoti}, \tool selects \texttt{ProcOpts(int, SubConf)} (line~J8), because this constructor initializes the \texttt{sub} field that the native side later reads (lines~N10--N13).

\noindent$\bullet$\textbf{ Sub-Argument Recursion and Assembly.} 
Once a path $\pi$ is selected, \tool recurses on each parameter type $\tau' \in \text{ParamTypes}(\pi)$ to obtain the corresponding sub-argument construction code $\mathit{ctor}(\tau')$ (lines 8--9); if any sub-argument cannot be constructed\zxy{, \tool discards $\pi$, backtracks to the path-selection loop to try a remaining candidate from $\mathit{ctorCands}$ (lines 5--10), and returns \textit{fail} once $\mathit{ctorCands}$ is exhausted (line 16)}. After all sub-arguments are obtained, \tool emits the reflective JNI API calls (e.g., \texttt{FindClass}, \texttt{GetMethodID}, and \texttt{NewObject}) that assemble the object at the current level (\zxy{lines 7--12}). 
Consider \texttt{ProcOpts} (lines~J5--J8 in~Fig.~\ref{figmoti}): after selecting $\pi = \texttt{ProcOpts(int, SubConf)}$ (line~J8) over \texttt{ProcOpts(int)} (line~J7), \tool recurses on its two formal parameter types, \texttt{int} and \texttt{SubConf}. While \texttt{int} terminates immediately as a directly synthesizable leaf type, the recursive step on the composite type \texttt{SubConf} (lines~J9--J11) selects $\texttt{SubConf(String, int)}$ (line~J11) and further recurses on its two leaf parameter types $\{\texttt{String}, \texttt{int}\}$, forming the Java-side expansion chain $\texttt{ProcOpts} \rightarrow \texttt{ProcOpts(int, SubConf)} \rightarrow \{\texttt{int}, \texttt{SubConf}\} \rightarrow \texttt{SubConf(String, int)} \rightarrow \{\texttt{String}, \texttt{int}\}$. This expansion chain is induced on the fly by the recursive calls over parameter types and tracked via the call stack and the agent's context, whereas only verified constructions are persisted in the harness memory(\S\ref{phase2.4harness-memory}) for subsequent reuse.

\noindent $\bullet$ \textbf{Local Check and Memoization.} Once assembly is complete, \tool performs a local check on $\textit{ctor}(\tau)$. First, it verifies that all referenced classes, constructors, fields, and JNI method signatures exist in the code artifacts. Second, it runs a smoke test to check whether the object can be instantiated, fields can be written, and the construction does not raise a Java exception (\zxy{line 13; otherwise, it backtracks to try the next candidate path}). Only $\textit{ctor}(\tau)$ that passes this check is added to the Memory in~\S\ref{phase2.4harness-memory} for reuse by subsequent JNI methods \zxy{(lines 14--15)}.

Overall, \tool produces, for every argument type along the call sequence, a piece of construction code \zxy{grounded in either bound context values or} the fuzzer byte stream. Together with the call sequence from~\S\ref{phase2.2call-sequence-extractor}, these snippets form the invocation context for the target JNI method.


\subsubsection{\textbf{Harness Memory Manager}}
\label{phase2.4harness-memory}
\zxy{While~\S\ref{phase2.2call-sequence-extractor} and~\S\ref{phase2.3arg-obj-constructor} can synthesize call sequences and argument constructions for individual JNI methods, processing hundreds of JNI targets within the same app introduces substantial redundancy. Specifically, the same composite argument types and precursor initialization routines frequently recur across related JNI methods; deriving them from scratch for each target repeatedly invokes retrieval tools, regenerates boilerplate code, and re-runs smoke tests. Because} harness synthesis is inherently a long-horizon task, \zxy{such repetition not only inflates LLM reasoning and verification overhead, but also fills the conversation history with redundant traces,} making it essential to manage the working memory of the LLM agent to mitigate the constraints of its limited context window and the degradation of long-range consistency.

To this end, \tool introduces a \textit{Harness Memory Manager} that maintains a global store of verified harness components. Whenever an intermediate component is validated (e.g., a call sequence that passes the resolution check, or an argument construction plan that passes the smoke test), the \textit{manager} commits it to the memory \zxy{alongside the metadata of the dependencies or object fields it initializes}. In subsequent rounds, when solving a new requirement, the \textit{manager} first queries the memory and reuses a matching component directly if a \zxy{compatible} one is already available\zxy{, namely one whose established state or initialized fields cover the requirements in $(S_m, A_m)$. If a reused component later fails downstream harness validation (\S\ref{phase3:validation}) due to stricter target-specific state constraints, the \textit{manager} retains the entry in global memory for other methods with simpler requirements, while bypassing it for the current target so that an alternative path can be synthesized.}

\subsection{Phase 3: Feedback-driven Harness Refiner}
\label{phase3}

With prerequisite call sequences and argument construction plans synthesized, \tool assembles them into runnable harnesses for fuzzers such as AFL++~\cite{leguevel_android_greybox_fuzzing_2023}, but two challenges arise. First, the JNI-aware harness should satisfy a chain of strict constraints (e.g., library loading, \texttt{JNIEnv} acquisition), and letting an LLM emit the entire harness in one shot opens a generation space wide enough for hallucinated, malformed code to slip through undetected. Second, harness validity cannot be judged by ``does not crash'' alone~\cite{ispoglou2020fuzzgen}: a harness that runs to completion may never enter the target JNI method's core logic, yet LLM agents readily mark such executions as successful, leaving downstream fuzzing trapped on shallow paths. \tool addresses these challenges with two designs.

\subsubsection{\textbf{App-Specific Harness Template}}
\tool constrains the LLM agent's generation space with a standardized JNI harness template that defines the overarching execution skeleton and the precise topological order in which individual functional sections are wired together, as illustrated in Fig.~\ref{fig:overview}; rather than synthesizing the entire harness from scratch in an unconstrained, error-prone manner, the agent is guided to populate designated slots with content derived from the previous two phases.

\subsubsection{\textbf{Validation-Guided Iterative Refinement}}
\label{phase3:validation}

\tool closes the synthesis loop through feedback-driven repair
and serves as an empirical validation boundary~\cite{sherman2025no} for the preceding
phases. After assembling a harness, \tool validates it with three
signals~\cite{sherman2025no}: \emph{(i)} compiler diagnostics, \emph{(ii)}
runtime exceptions, and \emph{(iii)} coverage
feedback, constraining the
LLM agent to perform targeted refinement rather than
unconstrained regeneration. \zxy{Specifically, using the target method's resolved library offset (\S\ref{phase1}) and Frida-based edge instrumentation, \tool executes the harness against a set of mutated trial inputs and considers it valid only if execution hits the target entry address and progressively covers new branches beyond the function prologue. This criterion overrides the agent's naive ``no-crash'' self-assessment and prevents shallow executions that repeatedly exit at initial guard checks (e.g., lines~N8 and~N13 in Fig.~\ref{figmoti}) from being misjudged as successful.}

\zxy{When validation fails,} \tool implements failure attribution
to identify the likely source of each failure, guiding the
\textit{Harness Refiner}: failures caused by malformed harness
code, such as compilation or linking errors, are handled in 
\S\ref{phase3} by re-synthesizing the affected template slots, whereas
failures indicating invalid argument state or missing
prerequisites, such as null-pointer dereferences, type
mismatches, uninitialized native handles, \zxy{or stagnant branch coverage at entry checks,} are routed back to
\S\ref{phase2}, which \zxy{bypasses any previously reused memory construction for the current target and} revises the corresponding
argument-construction plan or prerequisite call sequence. \zxy{Through this iterative loop, \S\ref{phase3} suppresses hallucinations and omissions accumulated across long-horizon reasoning, ensuring that every harness delivered to the fuzzer is grounded in dependencies and object states that have been verified on-device.}

\section{Implementation}
\label{sec:implementation}

    
\begin{figure}[t]
    \centering
    \includegraphics[trim=0 0 0 0, clip, width=\textwidth]{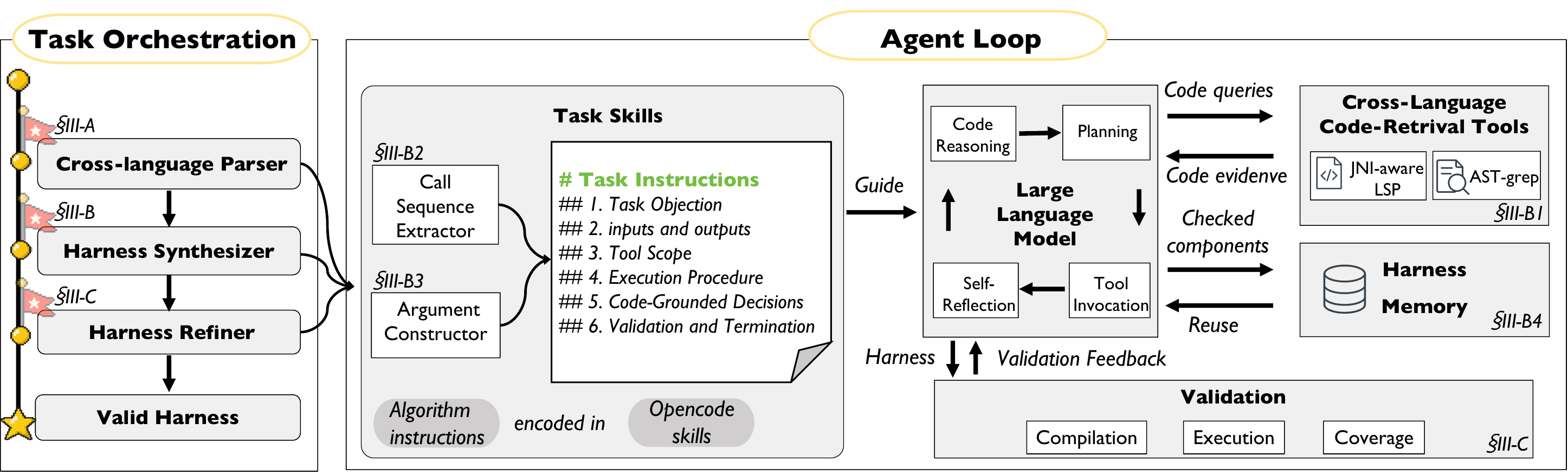}
    \caption{{Agentic implementation of \tool.}}
    \label{fig:Imp}
\Description{Agentic implementation of Janus connecting the agent workflow, code retrieval tools, persistent harness memory, compilation, and runtime execution feedback.}
\end{figure}

As shown in Fig.~4, \tool{} implements the proposed algorithms through task-specific Skills that guide the agent's cross-language information gathering, synthesis decisions, and validation. Specifically, these Skills encode the agentic workflow as procedural instructions within a customized extension of OpenCode~\cite{opencode_github}. Below, we describe four key aspects of how the agent follows these instructions.

\noindent $\bullet${\textbf{ Task Orchestration.}} \tool{} uses OpenCode agent to orchestrate the Cross-Language Parser (\S\ref{phase1}), the Harness Synthesizer (\S\ref{phase2}), and the Harness Refiner (\S\ref{phase3}). The project-level orchestration instructions in \texttt{src/AGENTS.md}~\cite{janus_anonymous_repo} specify the execution order of these stages and how refinement requests are routed. At each stage, the agent loads the corresponding Skill from \texttt{src/.opencode/}\cite{janus_anonymous_repo}. These Skills specify the overall workflow as concrete procedures, including JNI binding resolution, dependency-driven call sequence extraction, recursive argument construction, and continuous validation.

\noindent $\bullet${\textbf{ Algorithms Encoded as Skills.}} 
\tool's algorithms (described in \S\ref{phase2.2call-sequence-extractor} and \S\ref{phase2.3arg-obj-constructor}) explicitly specify task objectives, inputs and outputs, available tools, execution steps, decision criteria, and validation and termination conditions. The decision criteria distinguish explicit from implicit dependencies and directly synthesizable arguments from those requiring recursive construction. We encode our two key algorithms as OpenCode Skills to guide the agent's JNI-aware, app-specific execution loop. For Algorithm~\ref{algo1}, the Skill directs the agent to formulate queries based on unresolved explicit and implicit dependencies, identify prerequisite calls, extend the call sequence, and reassess the remaining dependencies. Once these dependencies are resolved, the agent performs a global review of the sequence. For Algorithm~\ref{algo:RecArgConstruction}, the Skill guides the agent through direct synthesis or memory reuse, construction-path selection, recursive sub-argument construction, and bottom-up assembly, with backtracking when a candidate construction fails. These instructions guide what the agent investigates, how it decomposes a synthesis task, and when it should continue, backtrack, or terminate.

\noindent $\bullet${\textbf{ {Code-Grounded Agent Loop.}}}
Within the algorithmic procedure encoded in each Skill, the agent combines code reasoning, planning, tool invocation, and self-reflection on intermediate results to recover app-specific JNI usage knowledge. This procedure involves identifying prerequisite method calls and determining how to construct complex Java arguments that satisfy implicit native-side expectations. Guided by multi-step planning, the agent retrieves relevant code through the JNI-aware LSP and AST-Grep interfaces (\S\ref{sec:phase2.1CodeRetriever}), which together provide the tool set $\mathcal{T}$ used by Algorithms~\ref{algo1} and~\ref{algo:RecArgConstruction}. The agent then applies code reasoning over the retrieved code facts for dependency classification and construction-path selection, while leveraging self-reflection to validate intermediate hypotheses and prune infeasible paths at the corresponding algorithmic steps.

\noindent $\bullet${\textbf{{ Checks, Memory, and Validation Feedback.}}}
Validation checks determine whether synthesized components are accepted for reuse: call sequences undergo global review, while argument-construction code undergoes local checks, including smoke tests. Accepted components are stored in the Harness Memory Manager (\S\ref{phase2.4harness-memory}) for subsequent reuse. Compiler diagnostics, runtime exceptions, and coverage measurements then provide feedback on the assembled harness. The failure-attribution mechanism (\S\ref{phase3}) uses this feedback to identify the likely source of failure and route refinement to the corresponding procedure. The agent revisits the relevant Skill to revise prerequisite calls, argument-construction code, or the affected harness template slots, connecting algorithm-guided synthesis with execution-based refinement.

\section{Effectiveness Evaluation}
\label{sec:Effectiveness}
Our evaluation of \tool addresses the following research questions (RQs).
\begin{itemize}
    \item \textbf{RQ1}: How effective is \tool in generating JNI-aware fuzzing harnesses for closed-source Android apps?
    \item \textbf{RQ2}: How does each component of \tool contribute to harness generation?
    \item \textbf{RQ3}: Can \tool discover previously unknown bugs in real-world Android apps?
\end{itemize}

\subsubsection*{\textbf{Experimental Environment}}
We conduct all experiments on a Mac Studio (M3 Ultra, 512 GB RAM). \tool{} exposes the structure-aware code retriever (\S\ref{sec:phase2.1CodeRetriever}) to the agent as tools under OpenCode's tool-calling convention. We adopt GLM-5 as the primary model throughout due to its cost-effectiveness~\cite{zeng2026glm}. Tests run on Android Emulator instances (API 31, 4 CPU cores, 6 GB RAM each), using Quarkslab's open-source Android runtime emulation scheme and Frida-driven AFL++ as the fuzzing engine~\cite{leguevel_android_greybox_fuzzing_2023}.

\subsubsection*{\textbf{Dataset}} We evaluate \tool on two complementary datasets that target different aspects of its effectiveness.

\noindent $\bullet$ \textbf{\textsc{Poirot}-44.}
We start from the public dataset released with
\textsc{Poirot}~\cite{di2025hercules}, which contains 103
popular real-world apps collected from Google Play. To
ensure reproducibility and a fair comparison with the
state-of-the-art baseline, we re-run \textsc{Poirot}'s
officially released raw Docker image without modification on this dataset; the image
successfully builds harnesses for only 44 of the 103 apps.
\zxy{We adopt these 44 apps, including 16 high-impact apps with
over 500M downloads (e.g., Google and YouTube Music),} as
our first dataset, 
comprising \zxy{21,386} JNI methods in total.
\textsc{Poirot}-44 is used in \textbf{RQ1} and
\textbf{RQ2} to evaluate harness-generation effectiveness
and the contribution of individual components, respectively.


\noindent $\bullet$ \textbf{Vendor-\zxy{20}.}
\label{sec:Vendor}
To further assess \tool's ability to discover unknown vulnerabilities in up-to-date, security-sensitive apps, we collect a second dataset (\textbf{Vendor-\zxy{20}}) by extracting \zxy{20} pre-installed apps from the latest flagship devices of \zxy{4} mainstream Android vendors (Xiaomi, Vivo, OnePlus, and OPPO; three to five per vendor; all running the February 5, 2026 security patch~\cite{android_security_bulletin_feb_2026}). All apps in Vendor-\zxy{20} share three properties: \emph{(i)}~they are vendor-preinstalled and span diverse functionalities including image processing, audio decoding, and font parsing; \emph{(ii)}~they rely heavily on native libraries for complex, error-prone low-level operations; and \emph{(iii)}~they typically run with \texttt{system} privileges, so any native-side memory vulnerability they harbor is significantly more impactful once exploited. Vendor-\zxy{20} is used exclusively in \textbf{RQ3} for in-the-wild vulnerability discovery.

%

\subsection{RQ1: Effectiveness of \tool's Harness Generation}
\label{sec:rq1}
\subsubsection{\textbf{Setup}}
We evaluate the effectiveness of \tool's harness generation on the
\textsc{Poirot}-44 dataset, comparing against the two most relevant Android JNI
fuzzing systems, Atlas++~\cite{atlasplusplus} and \textsc{Poirot}. Note that the original Atlas~\cite{zd} is not
publicly available; we therefore compare against Atlas++~\cite{atlasplusplus}, an open-source,
improved reimplementation released in April 2026.
Of the 21{,}386 JNI methods extracted from \textsc{Poirot}-44, 13{,}933 are excluded before harness generation, leaving 7{,}453 JNI targets for which \tool{} attempts harness generation. Specifically, we exclude JNI methods for which no call sites are identified in the host app's DEX bytecode, following established practice~\cite{di2025hercules}.

To ensure a fair comparison, all three tools are driven by the same
evaluation pipeline: identical APKs, emulator configurations, AFL++
instrumentation and seed corpora, and the same per-harness fuzzing budget.
During the harness-generation phase, each tool is first allotted a 30-minute timeout for its static-analysis stage, and is then given an additional 30-minute timeout per JNI target method for harness construction. \zxy{Following the \textsc{Poirot} evaluation protocol~\cite{di2025hercules}, each successfully built harness is allocated an AFL++ validation window of up to 4 minutes (approximately 29.6 days of compute time per run across 10,650 buildable harnesses). To reduce the non-determinism of the LLM, we fix the model temperature to 0 and average the results of \tool over three independent runs per JNI target method.}

\begin{table}[tbp]
 \centering
 \caption{Effectiveness comparison on \textsc{Poirot}-44 (best in \textbf{bold}).}
 \label{tab:rq1-perapp}
 \footnotesize
 \setlength{\tabcolsep}{1.6pt}
 \renewcommand{\arraystretch}{0.97}
 \begin{tabular*}{\linewidth}{@{\extracolsep{\fill}}l
 >{\columncolor{janusbg}}r >{\columncolor{janusbg}}r >{\columncolor{janusbg}}r >{\columncolor{januscr}}r
 >{\columncolor{atlasbg}}r >{\columncolor{atlasbg}}r >{\columncolor{atlasbg}}r >{\columncolor{atlascr}}r
 >{\columncolor{poirotbg}}r >{\columncolor{poirotbg}}r >{\columncolor{poirotbg}}r >{\columncolor{poirotcr}}r}
 \toprule
 &\multicolumn{4}{c}{\cellcolor{janushd}\textcolor{white}{\textbf{Janus}}}
 &\multicolumn{4}{c}{\cellcolor{atlashd}\textcolor{white}{\textbf{Atlas++}}}
 &\multicolumn{4}{c}{\cellcolor{poirothd}\textcolor{white}{\textbf{\textsc{Poirot}}}}\\
 App & \#Build & \#Run & Cov & \#Crashes & \#Build & \#Run & Cov & \#Crashes & \#Build & \#Run & Cov & \#Crashes \\
 \midrule
 UCMobile        & \textbf{578}  & \textbf{505}  & \textbf{22.6} & \textbf{2,920} & 93  & 4   & 4.3  & 2   & 84  & 84  & 4.8 & 33  \\
 amazon.mShop    & \textbf{394}  & \textbf{383}  & \textbf{23.1} & \textbf{293}  & 111 & 90  & 7.3  & 0   & 86  & 86  & 4.1 & 279 \\
 chrome          & \textbf{613}  & \textbf{582}  & \textbf{43.2} & \textbf{262}  & 80  & 18  & 18.4 & 0   & 67  & 67  & 12.4 & 250 \\
 apusapps        & \textbf{22}   & \textbf{21}   & \textbf{3.3}  & 0     & 8   & 2   & 0.4  & 0   & 4   & 4   & 1.1 & 0   \\
 billiards.pool  & \textbf{26}   & \textbf{21}   & \textbf{7.7}  & \textbf{4}    & \textbf{26}  & 5   & 2.3  & 0   & 7   & 6   & 2.6 & 0   \\
 bitstrips       & \textbf{86}   & \textbf{86}   & \textbf{8.9}  & \textbf{130}  & 5   & 0   & 0.0  & 0   & 32  & 32  & 2.4 & 124 \\
 cleanteam       & \textbf{23}   & \textbf{15}   & \textbf{21.2} & 0     & 1   & 0   & 0.0  & 0   & 3   & 1   & 0.8 & 0   \\
 skirt           & \textbf{36}   & \textbf{33}   & \textbf{5.4}  & \textbf{15}   & 25  & 5   & 2.3  & 0   & 10  & 10  & 2.9 & 0   \\
 larkplayer      & \textbf{19}   & \textbf{19}   & \textbf{11.1} & 0     & 10  & 8   & 2.9  & 0   & 5   & 5   & 1.7 & 0   \\
 evernote        & \textbf{20}   & \textbf{19}   & \textbf{2.0}  & \textbf{8}    & 8   & 3   & 1.5  & 7   & 5   & 2   & 1.3 & 0   \\
 gokids          & \textbf{19}   & 0    & 0.0  & 0     & 9   & 0   & 0.0  & 0   & 4   & 0   & 0.0 & 0   \\
 g.meetings      & \textbf{21}   & \textbf{20}   & \textbf{10.2} & \textbf{18}   & 5   & 4   & 3.0  & 0   & 4   & 4   & 2.2 & 17  \\
 g.photos        & \textbf{156}  & \textbf{110}  & \textbf{33.6} & \textbf{67}   & 20  & 10  & 19.1 & 0   & 21  & 21  & 8.7 & 64  \\
 yt.music        & \textbf{54}   & \textbf{50}   & \textbf{28.7} & \textbf{45}   & 2   & 0   & 0.0  & 0   & 15  & 15  & 3.6 & 43  \\
 g.dialer        & \textbf{12}   & \textbf{12}   & \textbf{24.0} & \textbf{25}   & 2   & 1   & 9.6  & 0   & 2   & 2   & 5.4 & 0   \\
 g.search        & \textbf{213}  & \textbf{189}  & \textbf{19.1} & 0     & 2   & 0   & 0.0  & 0   & 1   & 0   & 0.0 & 0   \\
 g.videos        & \textbf{73}   & \textbf{48}   & \textbf{11.7} & \textbf{5}    & 29  & 4   & 2.5  & 0   & 13  & 9   & 3.2 & 4   \\
 hicloud.clone   & \textbf{27}   & \textbf{27}   & \textbf{31.4} & \textbf{4}    & 17  & 17  & 3.2  & 3   & 8   & 8   & 2.1 & 0   \\
 huawei.maps     & \textbf{160}  & \textbf{160}  & \textbf{13.7} & \textbf{29}   & 59  & 44  & 2.8  & 28  & 39  & 39  & 2.4 & 0   \\
 imdb            & \textbf{146}  & \textbf{108}  & \textbf{10.7} & \textbf{10}   & 41  & 8   & 1.9  & 0   & 23  & 14  & 3.1 & 9   \\
 imo             & \textbf{128}  & \textbf{128}  & \textbf{12.2} & \textbf{15}   & 66  & 52  & 2.2  & 14  & 33  & 33  & 1.8 & 0   \\
 supervpn        & \textbf{4}    & \textbf{4}    & \textbf{15.6} & 0     & 2   & 0   & 0.0  & 0   & 1   & 1   & 0.9 & 0   \\
 parallel        & \textbf{31}   & \textbf{28}   & \textbf{4.6}  & \textbf{3}    & 13  & 4   & 0.5  & 0   & 5   & 5   & 1.2 & 0   \\
 linkedin        & \textbf{166}  & \textbf{166}  & \textbf{25.8} & \textbf{8}    & 46  & 35  & 2.2  & 7   & 23  & 23  & 1.6 & 0   \\
 skydrive        & \textbf{2,462} & \textbf{2,410} & \textbf{26.5} & \textbf{225}  & 743 & 323 & 2.0  & 214 & 396 & 396 & 3.4 & 0   \\
 brain.out       & \textbf{90}   & \textbf{90}   & \textbf{20.3} & \textbf{47}   & 32  & 31  & 2.1  & 45  & 27  & 27  & 1.9 & 0   \\
 talkinghank     & \textbf{14}   & \textbf{14}   & \textbf{5.9}  & \textbf{12}   & 4   & 2   & 0.7  & 0   & 2   & 2   & 0.6 & 0   \\
 tom             & \textbf{31}   & \textbf{15}   & \textbf{0.6}  & \textbf{14}   & 13  & 2   & 0.0  & 0   & 4   & 2   & 0.3 & 0   \\
 shealth         & \textbf{36}   & \textbf{36}   & \textbf{8.2}  & 0     & 25  & 11  & 2.2  & 0   & 12  & 12  & 2.5 & 0   \\
 easyMover       & \textbf{281}  & \textbf{281}  & \textbf{27.1} & \textbf{13}   & 72  & 71  & 2.7  & 12  & 40  & 40  & 2.3 & 0   \\
 printservice    & \textbf{40}   & \textbf{36}   & \textbf{20.9} & \textbf{19}   & 2   & 0   & 0.0  & 0   & 1   & 0   & 0.0 & 0   \\
 skype           & \textbf{715}  & \textbf{14}   & \textbf{11.8} & 0     & 353 & 3   & 2.0  & 0   & 213 & 6   & 2.8 & 0   \\
 skylight        & \textbf{67}   & \textbf{56}   & \textbf{69.7} & \textbf{57}   & 25  & 18  & 2.8  & 54  & 13  & 13  & 2.1 & 0   \\
 vectorunit      & 16   & \textbf{14}   & \textbf{2.7}  & 0     & \textbf{17}  & 0   & 0.0  & 0   & 7   & 1   & 0.7 & 0   \\
 midrop          & \textbf{24}   & \textbf{19}   & \textbf{38.1} & \textbf{40}   & 7   & 3   & 5.3  & 0   & 4   & 4   & 5.6 & 0   \\
 videoeditor     & \textbf{87}   & \textbf{26}   & \textbf{7.3}  & \textbf{7}    & 36  & 1   & 2.8  & 6   & 5   & 5   & 3.0 & 2   \\
 scaryteacher    & \textbf{36}   & \textbf{35}   & \textbf{18.3} & \textbf{74}   & 8   & 4   & 1.8  & 0   & 2   & 2   & 1.4 & 0   \\
 zeropage        & \textbf{58}   & \textbf{58}   & \textbf{1.8}  & \textbf{860}  & 8   & 4   & 0.9  & 0   & 3   & 3   & 0.7 & 0   \\
 superbros       & \textbf{23}   & \textbf{16}   & \textbf{1.2}  & \textbf{2}    & 9   & 1   & 0.7  & 0   & 4   & 4   & 0.8 & 1   \\
 turbovpn        & \textbf{54}   & \textbf{49}   & \textbf{12.2} & \textbf{15}   & 12  & 5   & 2.4  & 2   & 7   & 7   & 2.7 & 14  \\
 bubble.pop      & \textbf{41}   & \textbf{41}   & \textbf{36.4} & \textbf{25}   & 25  & 4   & 2.3  & 0   & 16  & 16  & 4.2 & 24  \\
 amazon.in       & \textbf{81}   & \textbf{81}   & \textbf{16.8} & \textbf{43}   & 20  & 16  & 2.8  & 10  & 18  & 18  & 3.1 & 41  \\
 pesam           & \textbf{34}   & \textbf{33}   & \textbf{7.6}  & \textbf{57}   & 22  & 4   & 3.0  & 0   & 4   & 4   & 2.4 & 0   \\
 firefox         & \textbf{40}   & \textbf{36}   & \textbf{20.9} & \textbf{19}   & 6   & 2   & 2.8  & 0   & 1   & 1   & 1.5 & 0   \\
 \midrule
 \rowcolor{totalbg}
 \textbf{Total} & \textbf{7,257} & \textbf{6,094} & \textbf{16.9} & \textbf{5,390}
 & 2,119 & 819  & 2.9  & 404
 & 1,274 & 1,034 & 2.6  & 905 \\
 \bottomrule
 \end{tabular*}
 \smallskip
 \noindent
 \begin{minipage}{\linewidth}
 \footnotesize
 \raggedright
 \textit{Note: Columns:} \textbf{\#Build}: harnesses that successfully compile, link, and package (no guarantee of valid execution). \textbf{\#Run}: harnesses that execute as valid fuzzing targets, i.e., initialize the runtime, load the target native library, consume inputs by executing the target function, and produce growing coverage. \textbf{Cov (\%)}: AFL++ bitmap coverage. Cov is measured only over the \#Run harnesses. \textbf{\#Crashes}: unique crashes from AFL's deduplication pipeline. The last row gives column totals (mean across apps for Cov). \tool{}'s improvements are statistically significant (paired Wilcoxon, $p<0.05$).
\end{minipage}
\end{table}

\subsubsection{\textbf{Result}}
The comparison results are shown in Table~\ref{tab:rq1-perapp}. \tool consistently outperforms both baselines across all metrics.
\zxy{Among the 7{,}453 attempted JNI targets, \tool constructs 7{,}257 buildable harnesses (97.4\%) and successfully executes 6{,}094 of them (84.0\% of the built harnesses). These counts correspond to 3.4$\times$/5.7$\times$ more buildable harnesses and 7.4$\times$/5.9$\times$ more runnable harnesses than Atlas++/\textsc{Poirot}, respectively.}
In terms of fuzzing effectiveness, \tool achieves a per-app mean bitmap coverage of 16.9\% (vs. 2.9\% for Atlas++ and 2.6\% for \textsc{Poirot}). After deduplication, \tool retains 5,390 crashes across 35 apps, compared to 404 (13 apps) and 905 (14 apps) for the two baselines. These gains are primarily attributable to
\tool's ability to capture complex, app-specific cross-language
dependencies and complex construction knowledge, enabling
it to generate more semantically meaningful fuzzing harnesses.

\noindent $\bullet$ \textbf{Failure Analysis.} Among the 7{,}257 harnesses generated by \tool, 6{,}094
satisfy the \#Run criterion (84.0\%); the remaining 1{,}163 fail
to execute successfully, indicating that the \textit{Harness Refiner} alone
is insufficient to guarantee that all harnesses are repairable. To
understand the root causes, we conduct a detailed manual analysis
of these failures. When the iteration budget is extended to two hours per harness,
637 of the 1{,}163 failed harnesses eventually succeed, reflecting
a cost--effectiveness trade-off rather than a fundamental
capability ceiling. The remaining failures fall into three
categories. \emph{(i)} Feedback in the emulated environment is
limited to coarse crash signals (e.g., SIGSEGV and its faulting
address), from which the LLM cannot always derive correct repair
guidance. \emph{(ii)} Some target libraries depend on external
runtime state, such as device sensors or system services, that
cannot be satisfied within the current emulated environment, and
such prerequisites cannot be eliminated through code-level
iterative repair alone. \emph{(iii)} Some target libraries contain
anti-emulation or integrity checks that trigger early exits in an
isolated harness environment, preventing stable execution
regardless of harness-level repairs.

\noindent $\bullet$ \textbf{ Crash Validation.} Since the analyzed libraries and apps are predominantly closed-source, manual analysis is highly labor-intensive. We select the 200 crashes with the highest coverage for manual inspection by three security researchers with over 3 years of experience (2--18 hours per crash), following a rigorous crash analysis procedure~\cite{di2025hercules}. We first
exclude 46 crashes that do not accept attacker-controlled inputs,
i.e., whose crash paths cannot be reached by external inputs.
\zxy{Of the remaining 154 crashes, we further discard 65 false positives, i.e.,
crashes not caused by defects in the target function itself. Their
root causes are threefold: (\emph{i}) incorrect invocation patterns inferred by
the LLM (12 crashes across 4 apps), (\emph{ii}) erroneous code information
produced by reverse-engineering tools due to obfuscation (23 crashes across
11 apps), and (\emph{iii}) benign anomalies, including exceptions actively
thrown by built-in exception-handling logic within the target method rather
than genuine defects (30 crashes across 9 apps).}
Among the 89 retained crashes, we confirm the
presence of defects in the corresponding functions. However, given
that the tested app versions are dated (2022), the underlying
vulnerabilities may have already been reported by others; we verify
this by checking the latest versions of the respective libraries and
confirm that all identified defects have since been patched.


\noindent $\bullet$ \textbf{Cost Analysis.} Averaged over three runs, each complete run of \tool consumes 2.57 billion tokens at an input-to-output ratio of 11.05:1, with
a context-cache hit rate of 94.7\%. The total API cost per run is
\$1{,}292.6, i.e., $\sim$\$0.212 per runnable harness,
demonstrating the economic practicality of \tool. 

\begin{mdframed}[
  skipabove=6pt,
  skipbelow=6pt,
  linewidth=1pt,
  linecolor=black,
  backgroundcolor=green!3,
  roundcorner=5pt
]
{\textbf{Answer to RQ1:}
\tool effectively generates fuzzing harnesses for native libraries of
Android apps. Over state-of-the-art baselines, \tool yields
$3.4\times$/$5.7\times$ more buildable and $7.4\times$/$5.9\times$ more
runnable harnesses, and uncovers 5{,}390 crashes.

}
\end{mdframed}

\subsection{RQ2: Ablation Study of the Analysis Passes}\label{Ganalysis}

\subsubsection{\textbf{Setup}}
RQ2 quantifies the contribution of each core component in \tool's Harness Synthesizer (\S\ref{phase2}). We reuse the RQ1 setup and metrics on \textsc{Poirot}-44 (44 apps, 21,386 JNI methods). 
We build four ablation variants, each disabling \tool's explicit workflow at a single stage and instead letting the LLM agent handle that stage unguided, isolating the contribution of each component relative to free-form LLM behavior. We also evaluated an out-of-the-box agent configuration (OpenCode+GLM-5 equipped with default file-inspection and shell tools, without \tool's synthesis workflows, structure-aware retrieval, component memory, or template-based harness validation). Under this setup, the agent failed to produce buildable harnesses from the decompiled APK artifacts; we therefore omit it from the quantitative comparison and focus on the controlled ablations above. 
The four ablation variants are:


\begin{itemize}
  \item \emph{\textbf{w/o-Retriever.}} Disables the structure-aware retriever
(\S\ref{sec:phase2.1CodeRetriever}); the agent explores
decompiled artifacts via textual \emph{grep} only, without
\tool's cross-language retriever.
  \item \emph{\textbf{w/o-CallSeq.}} Disables  the dependency-driven
  call-sequence extractor (\S\ref{phase2.2call-sequence-extractor}); the agent decides on
  its own how to set up the runtime state of the target method.
  \item \emph{\textbf{w/o-RecArg.}} Disables the recursive argument
  constructor (\S\ref{phase2.3arg-obj-constructor}); the agent synthesizes argument
  code freely.
  \item \emph{\textbf{w/o-Memory.}} Disables the memory
  (\S\ref{phase2.4harness-memory}); verified call sequences and constructors are no longer reused.
\end{itemize}

\begin{figure}[t]
\centering
\begin{tikzpicture}
\begin{axis}[
    ybar=2pt,
    bar width=16pt,
    width=\columnwidth,
    height=7.5cm,
    enlarge x limits=0.15,
    ymin=0, ymax=80,
    ylabel={Degradation vs.\ full \tool\ (\%)},
    ylabel style={font=\small},
    symbolic x coords={Build, Run, Cov, Crashes},
    xtick=data,
    xticklabels={\#Build, \#Run, Cov, \#Crashes},
    xticklabel style={font=\small},
    ytick={0,10,20,30,40,50,60,70,80},
    yticklabel style={font=\footnotesize},
    tick align=inside,
    grid=major,
    grid style={dashed, gray!30},
    legend style={
        at={(0.5,1.03)}, anchor=south,
        legend columns=4, font=\scriptsize,
        /tikz/every even column/.append style={column sep=2pt},
        draw=none,
    },
    nodes near coords,
    nodes near coords style={font=\tiny, rotate=90, anchor=west},
    every node near coord/.append style={/pgf/number format/fixed,
        /pgf/number format/precision=1},
]
\addplot[fill=blue!55, draw=blue!70!black] coordinates {
    (Build,32.0) (Run,38.4) (Cov,37.9) (Crashes,51.5)};
\addplot[fill=red!65, draw=red!80!black] coordinates {
    (Build,10.4) (Run,44.2) (Cov,50.4) (Crashes,62.2)};
\addplot[fill=orange!75, draw=orange!85!black] coordinates {
    (Build,5.7) (Run,25.9) (Cov,44.1) (Crashes,54.6)};
\addplot[fill=green!55!black, draw=green!40!black] coordinates {
    (Build,8.9) (Run,16.5) (Cov,17.9) (Crashes,30.2)};
\legend{w/o-Retriever, w/o-CallSeq, w/o-RecArg, w/o-Memory}
\end{axis}

\end{tikzpicture}
\caption{Component ablation: \emph{degradation} (\%) relative to full \tool.}
\label{fig:rq2-ablation}
\Description{Grouped bar chart of performance degradation relative to full Janus after removing Retriever, CallSeq, RecArg, or Memory. Metrics are buildable harnesses, runnable harnesses, coverage, and crashes.}
\end{figure}

\subsubsection{\textbf{Result}}
Disabling any component triggers a distinct regression (Fig.~\ref{fig:rq2-ablation}). These results show that each component contributes to the performance of the complete \tool configuration, with degradation patterns consistent with its intended role.

\noindent(\emph{i}) \textbf{Retriever underpins buildability.}
Restricted to textual \emph{grep}, the agent struggles to locate JNI entries, type definitions, and call contexts, so harnesses
are built on incomplete code understanding. This produces the largest \#Build drop ($-32.0\%$) and cascades to a substantial coverage drop (bitmap $16.9\%\!\to\!10.5\%$), establishing structure-aware retrieval as the foundation for all later synthesis steps. 
\noindent(\emph{ii}) \textbf{CallSeq bridges \emph{build}
and \emph{execution}.} Without prerequisite calls, harnesses
still build (\#Build $-10.4\%$, a modest drop) but crash
due to uninitialized runtime state before reaching the
target. The run/build ratio collapses from $84.0\%$ to
$52.3\%$, the lowest among all variants, confirming CallSeq
as the critical link from a \emph{buildable} harness to a
\emph{runnable} one.
\noindent(\emph{iii}) \textbf{RecArg governs exploration
depth.} Free-form arguments are typically type-correct but
semantically shallow (null objects, zero-length buffers).
Buildability is barely affected ($-5.7\%$), yet fuzzing
stalls at native input-validation prologues: coverage drops
to $9.4\%$. Argument legality and completeness are thus
prerequisites for penetrating into deep native logic. \noindent(\emph{iv}) \textbf{Memory sustains quality at scale.} Reuse is not required to get harnesses off the
ground, making this variant the mildest (Build $-8.9\%$,
Run $-16.5\%$). Its value is \emph{amplification}: without
it, similar methods are re-derived from scratch (multiplying
error chances), and high-coverage patterns cannot transfer
across related methods, shrinking the aggregate coverage
footprint.

\begin{mdframed}[
  skipabove=6pt,
  skipbelow=6pt,
  linewidth=1pt,
  linecolor=black,
  backgroundcolor=green!3,
  roundcorner=5pt
]
{\textbf{Answer to RQ2:}
LLM agents are not a silver bullet. \tool introduces key innovations: code retriever, call-sequence extractor, argument
constructor, and memory manager, all contributing to its superior
performance.
}
\end{mdframed}

\subsection{RQ3: Unknown Bug Discovery}\label{Ganalysis-rq3}

\subsubsection{\textbf{Setup}}

\begin{table}[tbp]
 \centering
 \caption{Effectiveness comparison on \textsc{Vendor-20} (best in \textbf{bold}).}
 \label{tab:rq1-vendor20}
 \footnotesize
 \setlength{\tabcolsep}{1.6pt}
 \renewcommand{\arraystretch}{0.97}
 \begin{tabular*}{\linewidth}{@{\extracolsep{\fill}}l
 >{\columncolor{janusbg}}r >{\columncolor{janusbg}}r >{\columncolor{janusbg}}r >{\columncolor{januscr}}r
 >{\columncolor{atlasbg}}r >{\columncolor{atlasbg}}r >{\columncolor{atlasbg}}r >{\columncolor{atlascr}}r
 >{\columncolor{poirotbg}}r >{\columncolor{poirotbg}}r >{\columncolor{poirotbg}}r >{\columncolor{poirotcr}}r}
 \toprule
 &\multicolumn{4}{c}{\cellcolor{janushd}\textcolor{white}{\textbf{Janus}}}
 &\multicolumn{4}{c}{\cellcolor{atlashd}\textcolor{white}{\textbf{Atlas++}}}
 &\multicolumn{4}{c}{\cellcolor{poirothd}\textcolor{white}{\textbf{\textsc{Poirot}}}}\\
 App & \#Build & \#Run & Cov & \#Crashes & \#Build & \#Run & Cov & \#Crashes & \#Build & \#Run & Cov & \#Crashes \\
 \midrule
 backuprestore     &    \textbf{46} &  \textbf{42} & \textbf{14.2} &   \textbf{4} &          11 &   3 &  2.1 &   0 &   7 &   7 & 1.8 &   0 \\
 bbk.theme         &   \textbf{213} &  \textbf{64} & \textbf{18.5} &   \textbf{5} &          64 &  22 &  2.7 &   4 &  32 &  18 & 2.0 &   0 \\
 copilot           &   \textbf{652} & \textbf{168} & \textbf{24.6} &  \textbf{28} &          88 &  11 &  9.8 &   0 &  79 &  38 & 6.4 &  22 \\
 extraphoto        &    \textbf{43} &  \textbf{31} & \textbf{22.4} &   \textbf{9} &           3 &   0 &  0.0 &   0 &  13 &   9 & 3.4 &   7 \\
 fileexplorer      &    \textbf{47} &  \textbf{46} & \textbf{20.0} &   \textbf{3} & \textbf{47} &  16 &  3.1 &   2 &   9 &   9 & 2.1 &   0 \\
 gallery3d         & \textbf{2,622} & \textbf{540} & \textbf{23.8} &  \textbf{42} &         710 &  78 &  2.0 &  18 & 365 & 104 & 3.2 &   0 \\
 heytap.themestore & \textbf{1,156} & \textbf{215} & \textbf{13.5} &  \textbf{16} &         280 &  18 &  1.7 &   0 & 154 &  42 & 2.3 &   0 \\
 market            &   \textbf{687} & \textbf{176} & \textbf{15.2} &  \textbf{14} &         152 &  34 &  2.6 &   2 &  98 &  29 & 2.9 &   9 \\
 mediaeditor       &   \textbf{523} & \textbf{194} & \textbf{26.4} &  \textbf{48} &         126 &  38 &  5.9 &   0 &  78 &  34 & 4.1 &  19 \\
 miui.gallery      &   \textbf{212} &  \textbf{82} & \textbf{25.1} &  \textbf{15} &          58 &  19 &  2.9 &   8 &  36 &  16 & 2.2 &   0 \\
 mms               &             21 &  \textbf{19} &  \textbf{6.4} &   \textbf{1} & \textbf{23} &   2 &  0.6 &   0 &   4 &   3 & 1.1 &   0 \\
 music             &   \textbf{687} & \textbf{218} & \textbf{21.3} &  \textbf{64} &         112 &   7 &  3.8 &   1 &  86 &  36 & 4.5 &   8 \\
 ocrscanner        &   \textbf{178} &  \textbf{68} & \textbf{11.2} &   \textbf{7} &          49 &   8 &  2.5 &   4 &  24 &  12 & 2.8 &   1 \\
 oplus.themestore  &   \textbf{287} &  \textbf{54} & \textbf{11.8} &   \textbf{3} &          96 &   5 &  1.8 &   0 &  58 &   8 & 2.4 &   0 \\
 pictorial         &   \textbf{175} &  \textbf{62} & \textbf{10.9} &   \textbf{8} &          14 &   0 &  0.0 &   0 &  20 &  11 & 1.5 &   0 \\
 reader            &     \textbf{8} &   \textbf{7} &  \textbf{9.5} &   \textbf{2} &  \textbf{8} &   0 &  0.0 &   0 &   1 &   0 & 0.0 &   0 \\
 scanner           &    \textbf{18} &  \textbf{18} & \textbf{17.6} &   \textbf{6} &           6 &   4 &  2.3 &   1 &   5 &   5 & 2.6 &   4 \\
 thememanager      &   \textbf{195} &  \textbf{64} & \textbf{18.9} &   \textbf{5} &          48 &  15 &  2.2 &   3 &  26 &  12 & 1.6 &   0 \\
 video             &    \textbf{89} &  \textbf{48} & \textbf{29.8} &  \textbf{26} &          16 &   8 & 11.2 &   0 &  18 &  11 & 6.2 &  14 \\
 wallpapers        &   \textbf{194} &  \textbf{58} &  \textbf{8.2} &            0 &          58 &  11 &  2.0 &   0 &  28 &  14 & 2.2 &   0 \\
 \midrule
 \rowcolor{totalbg}
 \textbf{Total}    & \textbf{8,053} & \textbf{2,174} & \textbf{17.5} & \textbf{306} &       1,969 & 299 &  3.0 &  43 & 1,141 & 418 & 2.8 &  84 \\
 \bottomrule
 \end{tabular*}
 \smallskip
 \noindent
 \begin{minipage}{\linewidth}
 \footnotesize
 \raggedright
 \textit{Note: Columns:} \textbf{\#Build}: harnesses that successfully compile, link, and package (no guarantee of valid execution). \textbf{\#Run}: harnesses that execute as valid fuzzing targets, i.e., initialize the runtime, load the target native library, consume inputs by executing the target function, and produce growing coverage. \textbf{Cov (\%)}: AFL++ bitmap coverage. Cov is measured only over the \#Run harnesses. \textbf{\#Crashes}: unique crashes from AFL's deduplication pipeline. The last row gives column totals (mean across apps for Cov). \textsc{Janus}'s improvements are statistically significant (paired Wilcoxon, $p < 0.05$).
\end{minipage}
\end{table}

RQ3 evaluates \tool's ability to discover unknown vulnerabilities in real-world, security-sensitive apps using the Vendor-20 dataset (described in \S\ref{sec:Vendor}). We extract 11{,}172 JNI methods from these apps and exclude 1{,}845 methods following the pre-filtering criteria in \S\ref{sec:rq1}, leaving 9{,}327 JNI targets. We subject these targets to the same harness-generation pipeline as in RQ1. For each runnable harness, we allocate a fuzzing budget of up to 12 hours with early termination upon coverage saturation (i.e., when no new branches are discovered within a 1-hour window~\cite{lipp2023green, liyanage2023reachable, liyanage2024extrapolating, boehme2020fuzzing, gao2023beyond}).


\begin{table}[htbp]
\centering
\begin{minipage}[t]{0.98\columnwidth}
\centering
\caption{Native bugs in vendor apps.}
\label{tab:vendor-native-bugs}
\small
\setlength{\tabcolsep}{4pt}
\renewcommand{\arraystretch}{1.18}
\setlength{\extrarowheight}{2pt}
\renewcommand{\tabularxcolumn}[1]{m{#1}}

\rowcolors{2}{black!6}{white}
\begin{tabularx}{\linewidth}{@{}
> {\hsize=0.60\hsize\linewidth=\hsize\centering\arraybackslash}X
> {\hsize=0.95\hsize\linewidth=\hsize\centering\arraybackslash}X
> {\hsize=1.15\hsize\linewidth=\hsize\centering\arraybackslash}X
> {\hsize=1.30\hsize\linewidth=\hsize\centering\arraybackslash}X
@{}}
\toprule
\textbf{Vendor} & \textbf{APP} & \textbf{Library} & \textbf{CWE Type} \\
\midrule

\textbf{Vivo}
& Theme Process
& \texttt{lib*\allowbreak{}Pro*.so}
& \bugtype{\texttt{CWE-787}: OOB write}{unchecked field write} \\

\textbf{Vivo}
& Theme Process
& \texttt{lib*\allowbreak{}Pro*.so}
& \bugtype{\texttt{CWE-502}: deserialization}{object-state corruption} \\

\textbf{Xiaomi}
& Theme Manager
& \texttt{lib*\_\allowbreak{}nati*.so}
& \bugtype{\texttt{CWE-787}: OOB write}{unchecked index write} \\

\textbf{OnePlus}
& Reader
& \texttt{libmo*.so}
& \bugtype{\texttt{CWE-190}: integer overflow}{unchecked size math} \\

\textbf{OPPO}
& Reader
& \texttt{libz*bi.so}
& \bugtype{\texttt{CWE-190}: integer overflow}{parser-state corruption} \\

\textbf{OPPO}
& Video Player
& \texttt{lib*media*.so}
& \bugtype{\texttt{CWE-787}: OOB write}{decode-state corruption} \\

\bottomrule
\end{tabularx}
\rowcolors{2}{}{}
\end{minipage}
\end{table}

\subsubsection{\textbf{Result}}
\tool successfully builds \zxy{8{,}053} harnesses from the \zxy{9{,}327} attempted JNI targets (\zxy{$86.3\%$}), of which \zxy{2{,}174} satisfy the \#Run criterion defined in RQ1 (\S\ref{sec:rq1}; \zxy{$27.0\%$} of the built harnesses, \zxy{outperforming Atlas++ and \textsc{Poirot} by $7.3\times$ and $5.2\times$}). \zxy{Compared with \textsc{Poirot-44}, the lower \#Run rate across all three tools on \textsc{Vendor-20} stems from two domain traits: many vendor-specific JNI methods depend on proprietary ROM services or hardware contexts unavailable in standard emulators (\S6.2.1), while numerous thin getter/setter JNI bridges in large apps (e.g., \texttt{gallery3d}) yield flat coverage and are conservatively excluded by our strict \#Run check (\S3.3.2).} Across all evaluated APKs, \tool produces \zxy{2{,}535} raw crashing inputs in total. We first deduplicate them based on the crash signal, the target JNI function, the crashing stage, the faulting Program Counter, and the native backtrace, yielding \zxy{306} distinct crash groups. We then manually triage these groups, filtering out crashes that are unreachable from attacker-controllable production inputs. Finally, root-cause and reachability analyses confirm 6 vulnerabilities, as shown in Table~\ref{tab:vendor-native-bugs}. In our evaluation, neither Atlas++ nor \textsc{Poirot} detected any of the six vulnerabilities identified by \tool.

We have responsibly disclosed these vulnerabilities to the corresponding vendors, accompanied by manually validated bug reports, harness artifacts, and crash samples generated by \tool~\cite{janus_anonymous_repo}. At the time of submission, one OPPO vulnerability (internal tracking ID: OSRC-2026-2112960957058781184) has been confirmed by the vendor as a zero-day vulnerability with a bug bounty awarded. Additionally, one OPPO vulnerability and one OnePlus vulnerability have been acknowledged (internal tracking IDs: OSRC-2026-2115741462279884800 and OSRC-2026-2115777492253020160) and are currently under vendor investigation. The remaining three reports are still awaiting triage due to vendor backlogs.

\begingroup

\subsubsection*{\textbf{Case Study: Heap Out-of-Bounds Write in an Embedded Video Player}}


\tool{} uncovered a heap out-of-bounds write in the bundled native media decoder (\texttt{lib*media*.so}, invoked via its JNI bridge \texttt{lib*JNI*.so}) of a preinstalled video player. When decoding a crafted AVI file containing a video stream, the progressive 2-D reconstruction path validates the header-declared \texttt{lowpass\_width} only against its internal band allocation, given by $\texttt{ALIGN(coded\_width, 64)}/2$, failing to check it against the actual output image-plane width and buffer size. During horizontal filtering, the decoder writes $2 \times \texttt{lowpass\_width}$ 16-bit samples per row along the plane line stride; for non-block-aligned coded widths, each row spills past the row boundary and the final row writes decoded pixel data beyond the output plane allocation into adjacent heap memory. This corrupts live heap objects and allocator metadata (e.g., Skia rendering pointers, frame-data pointers, or Bionic tagged pointers) and causes fatal process crashes (\texttt{SIGSEGV} during rendering or \texttt{SIGABRT} during memory deallocation), which we confirmed on a physical OPPO device using multi-frame samples. In practice, an attacker can trigger this vulnerability remotely by delivering a crafted video file through messaging, downloads, or web links, or locally without permissions by passing a media URI to the player's exported, permission-free \texttt{ACTION\_VIEW} playback activity.


\endgroup

\subsubsection*{\textbf{Case Study: Out-of-Bounds Access in Vendor Theme Manager}}

To illustrate \tool in a real deployment, we examine an out-of-bounds write that \tool surfaced in a native rendering library (lib*\_\allowbreak{}nati*.so) of a privileged theme-management app (Theme Manager) on Xiaomi flagship devices. The triggering harness first issues a sequence of precursor calls to establish the renderer's uniform-buffer state, then mutates the two index arguments passed to the vulnerable function while enforcing the array-length constraints required by the JNI byte-array interface.

The app runs with system privileges and renders user-supplied visual scenes packaged in a custom ZIP-based archive that bundles scene descriptors and assets. The native function derives its write target by multiplying a caller-supplied instance index by a fixed byte stride, but never bounds-checks the index; an oversized index drives the write past the allocated buffer into adjacent heap memory. Since both the write offset (the index) and the written value (a descriptor attribute) are attacker-controlled, the defect constitutes a controllable relative heap-write primitive, remotely reachable by luring a victim into applying a crafted scene file.
\zxy{We also reproduced the crash associated with this vulnerability on a physical Xiaomi device.}

\begin{mdframed}[
  skipabove=6pt,
  skipbelow=6pt,
  linewidth=1pt,
  linecolor=black,
  backgroundcolor=green!3,
  roundcorner=5pt
]
\textbf{Answer to RQ3:} \tool effectively discovers previously unknown vulnerabilities in real-world Android apps, confirming 6 vulnerabilities across 4 major vendors.
\end{mdframed}

\section{Discussion}\label{sec:discussion}

\subsection{Choice of Base Model}\label{sec:base-model}
Choosing a backbone for \tool requires balancing long-horizon tool use, harness quality, inference cost, and reproducibility. We tested open-weight models, including DeepSeek-V3.2~\cite{liu2025deepseek}, Kimi-K2.5~\cite{team2026kimi}, Qwen-3.5~\cite{team2026qwen3}, and GLM-5~\cite{zeng2026glm}, as well as proprietary models, including GPT-5.2~\cite{openai_gpt_5_2_2025} and Claude Opus 4.5~\cite{anthropic_claude_opus_4_5_2025}.

To substantiate our choice, we conducted a controlled experiment that varied \emph{only the underlying LLM}, while keeping the \tool workflow, tools, budget, and evaluation pipeline identical. We randomly sampled five apps from \textsc{Poirot}-44, covering 225 JNI methods in total. Table~\ref{tab:base-model-comparison} reports buildable harnesses (\#Build) and runnable harnesses (\#Run), using the same definitions as in RQ1.

\begin{table}[t]
\centering
\begin{minipage}[t]{0.98\columnwidth}
\centering
\caption{Backbone model comparison on five randomly sampled apps from \textsc{Poirot}-44 (225 JNI methods).}
\label{tab:base-model-comparison}
\small
\setlength{\tabcolsep}{1.5pt}
\renewcommand{\arraystretch}{1.12}
\rowcolors{3}{white}{black!6}
\newcommand{\janusmodelhead}[2]{%
  \ensuremath{\vcenter{\hbox{\shortstack[c]{\textbf{#1}\\\textbf{#2}}}}}%
}
\begin{tabularx}{\linewidth}{@{}cc*{5}{>{\centering\arraybackslash}X}@{}}
\toprule
\rowcolor{white}
\multicolumn{7}{c}{\textbf{(a) Buildable harnesses (\#Build)}} \\
\midrule
\rowcolor{black!6}
\textbf{Package} & \janusmodelhead{DeepSeek-}{V3.2} & \janusmodelhead{Kimi-}{K2.5}
& \janusmodelhead{Qwen-}{3.5} & \textbf{GLM-5} & \janusmodelhead{Claude}{Opus 4.5} & \textbf{GPT-5.2} \\
\midrule
\texttt{vectorunit} & 2 & 8 & 5 & 16 & 18 & 16 \\
\texttt{evernote} & 1 & 4 & 3 & 20 & 20 & 20 \\
\texttt{superbros} & 0 & 7 & 12 & 23 & 26 & 22 \\
\texttt{billiards.pool} & 0 & 3 & 13 & 26 & 26 & 30 \\
\texttt{talkinghank} & 3 & 9 & 7 & 14 & 14 & 14 \\
\midrule
\rowcolor{white}
\textbf{Total} & 6 & 31 & 40 & 99 & 104 & 102 \\
\midrule
\rowcolor{white}
\multicolumn{7}{c}{\textbf{(b) Runnable harnesses (\#Run)}} \\
\midrule
\rowcolor{black!6}
\textbf{Package} & \janusmodelhead{DeepSeek-}{V3.2} & \janusmodelhead{Kimi-}{K2.5}
& \janusmodelhead{Qwen-}{3.5} & \textbf{GLM-5} & \janusmodelhead{Claude}{Opus 4.5} & \textbf{GPT-5.2} \\
\midrule
\texttt{vectorunit} & 0 & 3 & 0 & 14 & 18 & 15 \\
\texttt{evernote} & 0 & 1 & 1 & 19 & 20 & 19 \\
\texttt{superbros} & 0 & 3 & 10 & 16 & 18 & 15 \\
\texttt{billiards.pool} & 0 & 0 & 7 & 21 & 23 & 18 \\
\texttt{talkinghank} & 0 & 2 & 1 & 14 & 14 & 14 \\
\midrule
\rowcolor{white}
\textbf{Total} & 0 & 9 & 19 & 84 & 93 & 81 \\
\bottomrule
\end{tabularx}
\rowcolors{2}{}{}
\end{minipage}
\end{table}

\noindent$\bullet$\textbf{ Sustaining the Agentic Workflow.}
In this evaluation, DeepSeek-V3.2, Kimi-K2.5, and Qwen-3.5 synthesized 6, 31, and 40 buildable harnesses, respectively, but only 0, 9, and 19 of them were runnable. When executing long-horizon reasoning, these models frequently violated the tool-calling protocol or terminated prematurely before resolving runtime JNI constraints. These results indicate that, across the tested models and apps, generating syntactically buildable code alone does not guarantee that a backbone LLM can reliably sustain \tool's multi-step agentic workflow.

\noindent$\bullet$\textbf{ Harness Quality Relative to Frontier Models.}
GLM-5 produced 99 buildable and 84 runnable harnesses, compared with 104 and 93 for Claude Opus 4.5, the strongest backbone by both aggregate counts in this subset. Thus, GLM-5 recovered 95.2\% ($99/104$) of Opus 4.5's buildable harness count and 90.3\% ($84/93$) of its runnable harness count. GPT-5.2 produced 102 buildable and 81 runnable harnesses. These results show that GLM-5 approaches the frontier models in harness generation on this subset, while mitigating the severe workflow failures observed with the other tested open-weight backbones.

\noindent$\bullet$\textbf{ Cost and Reproducibility Trade-Off.}
The broader cost comparison concerns the full \textsc{Poirot}-44 workload rather than this five-app experiment: the Opus 4.5 cost is \$7{,}225.8, approximately 5.6$\times$ the GLM-5 API cost of \$1{,}292.6. As reported in RQ1, GLM-5 costs approximately \$0.21 per runnable harness. Its open weights also facilitate independent deployment and inspection, whereas the closed nature of the proprietary models limits reproducibility. Together, its near-frontier harness counts on the sampled apps, robust instruction following, open weights, and low cost justify GLM-5 as \tool's default backbone.

We expect continued improvements in open-weight models to further narrow the observed gap and improve cost-effectiveness, making open, reproducible, and affordable backbones increasingly attractive for agentic harness synthesis. 



\subsection{Limitations and Threats to Validity}\label{sec:limitations} 
\subsubsection{\textbf{Limitations}}
\tool's primary limitations stem from the following aspects, which also highlight potential avenues for future research. 

\noindent$\bullet$\textbf{ Concurrent and Asynchronous Dependencies.} \tool currently focuses on resolving explicit and implicit dependencies to reconstruct the JNI execution context, but is inherently based on a linear execution model. However, Android apps frequently involve complex asynchronous mechanisms (e.g., coroutines) and concurrent mechanisms (e.g., multi-threading). \tool may fail to satisfy state dependencies established by asynchronous calls, and may miss concurrency-related vulnerabilities in highly parallelized native libraries. 

\noindent$\bullet$\textbf{ Framework and System Service Interactions.} \tool currently focuses on modeling the JNI call paths within the app itself, but does not account for deep interactions with the Android (e.g., Binder). Such interactions require complex real-time runtime contexts; consequently, for native libraries that rely heavily on such system interactions, \tool may fail to generate harnesses.

\subsubsection{\textbf{Threats to Validity}}
The primary threat to validity stems from the generalizability of our evaluation datasets. \textsc{Poirot}-44 represents a traditional category of Android commercial apps that run with normal user-level permissions. Evaluating solely on such apps may limit the generalizability of our findings to other apps that differ in terms of provenance, permission levels, and native library characteristics. To mitigate this threat, we introduce Vendor-20, a distinct category of vendor pre-installed apps granted system-level permissions, which exhibits different library traits and access privileges. While these two datasets enrich the diversity of our evaluation, other categories of Android apps may remain uncovered; we leave a broader evaluation across a wider spectrum of app types to future work.

%

\section{Related Work}
\subsection{Automated Harness Generation for Libraries}
To reduce the cost of manually writing fuzzing harnesses, a series
of recent works have proposed methods for automatically synthesizing
harnesses~\cite{sherman2026snakecharmer,sherman2025no,jeong2023utopia,green2022graphfuzz,zhang2021apicraft,zhang2023daisy,jung2021winnie,babic2019fudge,chen2023hopper,ispoglou2020fuzzgen}. Specifically, 
OGHarn~\cite{sherman2025no}
splices candidate C-API harnesses based on header files and produces
harnesses through filtering and mutation. Some efforts also target
cross-language harness synthesis, such as
SnakeCharmer~\cite{sherman2026snakecharmer}, which recovers interface types and
exception semantics for Python~$\leftrightarrow$~C/C\texttt{++}
hybrid libraries and generates harnesses accordingly. More recently,
studies have begun to delegate harness synthesis tasks to LLMs and
LLM-driven agents~\cite{zeng2025pbfuzz,zhang2024effective,yang2025harnessagent,zoustepstone,xu2025ckgfuzzer,lyu2024prompt,xia2024fuzz4all,deng2023large,meng2024large,yang2025hybrid,chen2025elfuzz,kim2025atlantis,sheng2025all,liu2024testing}, demonstrating impressive results in
extracting API usage knowledge and generating harnesses. While \textbf{source-level LLM pipelines} (e.g., LLM4FDG~\cite{zhang2024effective} and CKGFuzzer~\cite{xu2025ckgfuzzer}) and \textbf{agentic approaches} such as HarnessAgent~\cite{yang2025harnessagent} effectively synthesize harnesses for open-source C/C\texttt{++} projects by querying source trees (e.g., headers, type definitions, and call graphs) and routing compilation errors, \textbf{they do not address two cross-language challenges inherent to closed-source Android native libraries}. (\emph{i}) unlike C/C\texttt{++} targets whose API dependencies and state transitions reside within a unified source codebase, Android native libraries are stripped \texttt{.so} binaries driven by a managed Java host; valid execution requires recovering app-specific prerequisite calls and tracking explicit value and implicit state bindings across the Java-to-native boundary. (\emph{ii}) whereas C/C\texttt{++} harnesses instantiate argument types defined directly in native headers, JNI targets accept opaque object references (\texttt{jobject}) whose valid construction paths reside in decompiled Java code while their actual field consumption occurs in native code. Synthesizing valid arguments therefore requires reconciling native-side field-access demands with app-side Java construction paths to recursively assemble composite objects.

\subsection{Android/JNI Vulnerability Detection}
Existing Android vulnerability detection research largely focuses on
the Java/Kotlin layer, with limited coverage of native-code security
risks~\cite{mao2025nass,acar2016sok,xue2018ndroid,wei2018jn,almanee2021too,samhi2022jucify,borzacchiello2022reach,zd,di2025hercules,auer2023android,kalysch2020template,maji2012empirical,kai2015fuzzing}. Cross-language static analyses such as
NativeSummary~\cite{wang2024nativesummary} extract JNI summaries from native binaries for
bytecode analyzers, but target information-flow tracking rather than
memory corruption. In terms of Android fuzzing studies: FANS~\cite{liu2020fans} synthesizes
harnesses from Android system-service source code but assumes source
availability; JNIFuzzer~\cite{rizzo2020static} fuzzes JNI methods yet lacks
Java-side context modeling. Atlas~\cite{zd} and \textsc{Poirot}~\cite{di2025hercules} are the closest systems for closed-source native libraries; both infer app-specific JNI usage through static program analysis. Both remain constrained by the precision-scalability trade-off, and neither recovers construction knowledge for complex composite objects in real-world applications.


\subsection{Cross-Language Analysis of Android Applications}
\label{sec:rw-cross-language-analysis}
Prior research~\cite{wei2018jn,xue2018ndroid,samhi2022jucify,borzacchiello2022reach,wang2024nativesummary} has explored static cross-language program analysis across the Java and native boundaries, alongside general-purpose JNI analysis frameworks such as AXA~\cite{roth2024axa} and Confluence~\cite{prakash2025confluence}. For instance, JN-SAF~\cite{wei2018jn} and NDroid~\cite{xue2018ndroid} track inter-language data and taint flows, JuCify~\cite{samhi2022jucify} unifies Java and native intermediate representations to construct cross-language call graphs (CGs), and NativeSummary~\cite{wang2024nativesummary} extracts native binary summaries to assist bytecode analyzers.

However, these static cross-language analyzers cannot be directly applied to synthesize fuzzing harnesses for closed-source Android native libraries due to three fundamental distinctions. (\emph{i}) \textbf{Recovering cross-language call relations does not equate to executable harness synthesis}. Existing tools primarily focus on structural reachability~\cite{borzacchiello2022reach} or information-flow tracking~\cite{wei2018jn,wang2024nativesummary}, rather than on synthesizing standalone C/C\texttt{++} driver code that reconstructs implicit native-side runtime states or recursively instantiates complex composite Java arguments via reflective JNI APIs. (\emph{ii}) \textbf{General-purpose Java-native analysis frameworks face ecosystem incompatibilities on Android}. Frameworks such as AXA rely on Java bytecode analyzers (e.g., OPAL) tailored for standard Java bytecode rather than Android DEX bytecode and lifecycle-driven app components, rendering them unsuitable for analyzing commercial Android APKs. (\emph{iii}) \textbf{CG-based approaches suffer from a severe precision-scalability trade-off on modern apps}. Heavyweight cross-language CG construction and symbolic analysis (e.g., JuCify~\cite{samhi2022jucify}, Atlas~\cite{zd}) frequently time out or yield empty results on large-scale, obfuscated apps. Conversely, lightweight Java-side CG approximations (e.g., \textsc{Poirot}~\cite{di2025hercules}) miss native-side state transitions and cross-callback initialization sequences (such as global states initialized in lifecycle entry points outside the target's direct call chain), omitting critical setup and argument construction and thereby inducing false-positive crashes from corrupted runtime states.

To mitigate the limitations of conventional cross-language program analysis in recovering the JNI usage knowledge required for \textbf{executable harness generation}, \tool introduces depen\-dency-driven call-sequence extraction to reconstruct the app-specific prerequisite context needed for valid JNI execution. It further exploits semantic relationships among parameters to recursively construct complex composite arguments, enabling the generated harnesses to effectively exercise closed-source Android native libraries.

\section{Conclusion}
In this paper, we propose \tool, the first agentic approach to synthesizing JNI-aware fuzzing harnesses for Android native libraries. Grounded in structure-aware cross-language code retrieval, \tool iteratively resolves both explicit and implicit call dependencies to reconstruct app-specific invocation contexts, and recursively assembles complex composite argument objects in a bottom-up manner, supported by harness memory and validation-guided iterative refinement. Extensive evaluation on real-world commercial and vendor pre-installed apps demonstrates that \tool outperforms state-of-the-art baselines, yielding more valid harnesses ($7.4\times$/$5.9\times$ more runnable harnesses) and uncovering previously unknown vulnerabilities. Our work lays the foundation for agentic cross-language fuzzing harness generation in the Android ecosystem. Beyond standalone harness synthesis, \tool sheds light on downstream research fields for Android native code security. Many current and future Android cross-language security vetting and dynamic analysis initiatives can be enhanced by incorporating \tool's insights.

\section{Data Availability}

To facilitate reproducibility and foster future research, we publicly release our prototype and artifacts~\cite{janus_anonymous_repo}.
 
\section*{Acknowledgments}
This work was supported by the National Key R\&D Program of China (Grant No.2024YFE0203800), the National Natural Science Foundation of China (No. 62472309), and the grant from Zhongguancun Academy (Grant No. 20240302).


\bibliography{sample-base} 

\end{document}